%% file: main.tex
\documentclass[12pt]{article}
\usepackage[utf8]{inputenc}
\usepackage{graphicx}
\usepackage{lscape}
\usepackage{url}
\usepackage{marvosym}
\usepackage{abstract}
\usepackage{amsmath}
\usepackage{natbib}
\usepackage{soul,color}
\usepackage[toc,page]{appendix}
\usepackage{adjustbox} 
\usepackage{hyperref}
\newcommand{\footremember}[2]{%
  \footnote{#2}
  \newcounter{#1}
  \setcounter{#1}{\value{footnote}}%
}
\newcommand{\footrecall}[1]{%
  \footnotemark[\value{#1}]%
} 

\usepackage{hyperref}

\providecommand{\keywords}[1]
{
 \small	
 \textbf{\textit{Keywords---}} #1
}

\usepackage{geometry}
\title{Monitoring COVID-19-induced gender differences in teleworking rates using Mobile Network Data}

\author{Sara Grubanov-Boskovic  \footnote{Corresponding author, Email: \href{mailto:Sara.GRUBANOV-BOSKOVIC@ec.europa.eu}{Sara.GRUBANOV-BOSKOVIC@ec.europa.eu} } \footremember{aldo}{European Commission, 
              Joint Research Centre,
              Ispra (VA), Italy}  \and
    Spyridon Spyratos\footrecall{aldo} \and
    Stefano Maria Iacus\footrecall{aldo}   \and
    Umberto Minora\footrecall{aldo} \and
    Francesco Sermi\footrecall{aldo} \and
}

\date{November 2021}

\begin{document}

\maketitle

\begin{abstract}

The COVID-19 pandemic has created a sudden need for a wider uptake of home-based telework as means of sustaining the production. Generally, teleworking arrangements  impacts directly worker’s efficiency and motivation. The direction of this impact, however, depends on the balance between positive effects of teleworking (e.g. increased flexibility and autonomy) and its downsides (e.g. blurring boundaries between private and work life). Moreover, these effects of teleworking can be amplified in case of vulnerable groups of workers, such as women. The first step in understanding the implications of teleworking on women is to have timely information on the extent of teleworking by age and gender. In the absence of timely official statistics, in this paper we propose a method for nowcasting the teleworking trends by age and gender for 20 Italian regions using mobile network operators (MNO) data. The method is developed and validated using MNO data together with the Italian quarterly Labour Force Survey. Our results confirm that the MNO data have the potential to be used as a tool for monitoring gender and age differences in teleworking patterns. This tool becomes even more important today as it could support the adequate gender mainstreaming in the ‘Next Generation EU’ recovery plan and help to manage related social impacts of COVID-19 through policymaking.

\end{abstract}

\section{Introduction} 

The COVID-19 pandemic has been a determinative factor in accelerating the uptake of teleworking arrangements around the World. This rapid, large-scale adoption of teleworking has brought new opportunities, but also challenges. Ultimately, how families, firms and society at large manage to harness the opportunities of teleworking, while minimizing its downsides, will determine its short- and long-term economic and social impacts. 
On the labour market, the teleworking has a direct impact on the worker’s efficiency and motivation, which in turn influence firm productivity. As the recent literature reviews have pointed out \citep{samek2021,oecd2020}, the worker's efficiency and motivation depend on the balance between positive effects of teleworking (better work-life balance, increased flexibility and autonomy, reduction in commuting time, etc.) and its downsides (blurring boundaries between private and work life, overtime work, inappropriate working environment at home, social isolation, etc.).
How teleworking is being implemented and managed at family, firm and society level does not carry only economic implications – e.g. through job satisfaction and firm productivity or through firm savings – but important social spillovers as well, for example on gender equality \citep{oecd2020}. 

\paragraph{How teleworking can affect gender gaps?}\mbox{} 
\\
Since the second half of the 20th century, Europe has been transitioning from a traditional family model of ‘male breadwinner - female care provider’ toward the model of ‘dual earner-carer’ family, based on the principle of equal participation of both women and men in paid and unpaid work \citep{Esping-Andersen2015,Esping-Andersen2013, hobson_2004}. This transition has led to an increase in the female labour market participation and employment over time, nonetheless, when compared to men, important gender gaps still remain in terms of both paid work (employment, wages, occupational segregation) and unpaid work (domestic and childcare work) \citep{Grubanov2020,Blau2017}.

In the pre-pandemic context, the literature has highlighted the relevance of flexible working arrangements, such as telework, in supporting female labour market participation. According to the 2015 European Working Conditions Survey data, women with children had higher probability of engaging in telework as means of maintaining life-work balance, in comparison to other categories of workers \citep{Eurofound}. Moreover, women with children who had the possibility to telework had lower probability of reducing their working hours after childbirth \citep{Chung2018} and also faced lower employment barriers especially in higher-paid jobs \citep{Fuller2019} . 

As the COVID-19 pandemic unfolded, its adverse implications for gender equality started penetrating not only the dimension of paid work, but also the area of unpaid (domestic and childcare) work.

Considering the dimension of paid work, numerous studies have pointed to adverse effects of COVID-19 pandemics on the employment and income of women, as a result of higher share of female employment in sectors which were strongly hit by the crisis \citep{Dang2021,Farre2020,alon2020,FullerQian2021,KRISTAL2020100520}. Nonetheless, the evidence is still not conclusive. For example, \cite{ADAMSPRASSL2020104245} find that women have had higher probability of losing their job during COVID-19 in the USA and the UK, but not in Germany. Likewise, \cite{Casarico2020} in case of Italy, and \cite{hupkauPertongolo2020} in case of the UK, do not find an overall gender-differentiated effect of the pandemic on job separations. These differing results can be explained both in terms of methodological differences, but also in terms of cross-country differences in policies and sectoral distribution of female workers. \cite{Casarico2020}, for example, argue that country-specific policies protecting more vulnerable workers, such as women, but also larger female employment in occupations with teleworkable tasks in certain regions might have contributed shielding women, at least temporarily, from adverse effects of the crisis. While teleworking has sustained the female employment to a certain degree \citep{Farre2020}, at the same time, some initial evidence suggest that teleworking during the pandemics might have increased the gender pay gap. For example, \cite{Bonacini2021} find that  the increase in teleworking arrangements in Italy has been associated with an increase in the average labour income only for men, and not for women. 

The growing literature has also been analysing how COVID-19 pandemics has affected gender equality in terms of equal division of unpaid labour between men and women, given the large increase in childcare and housework generated by lockdowns, school closures and the inability to outsource. The current evidence suggests that men have indeed increased their participation in unpaid work in comparison to the pre-pandemic period, however, the larger chunk of unpaid work has still fell on women, especially working mothers \citep{Farre2020,Chung2021,Hipp2021,Dunatchik2021,HupkayPetrongolo2020}. However, in cases in which both parents teleworked the gender gap in unpaid labour remained largely the unvaried \citep{Dunatchik2021}. 

Overall, the evidence shows that expansion of telework for all workers could lead to larger involvement of men in unpaid work, which in turn could contribute eroding social norms underpinning the gendered division of housework and caregiving tasks, and ultimately improve the labour market outcomes of women \citep{alon2020,Petts2021,samek2021}. At the same time, however, the literature points out that expanding the flexible working alone (e.g. telework) is not a sufficient enough condition for narrowing the gender gap. Specifically, several authors (\cite{Chung2021,rubery2020covid}) raise concerns of a risk of regressing back toward the ‘male breadwinner - female care provider’ if the expansion of flexible working arrangement is not accompanied by policies that are effective in influencing societal norms around gender roles. In case of Europe, for example, \cite{rubery2020covid} call for gender mainstreaming in the ‘Next Generation EU’ recovery programme as means of closing the gender gaps in both paid and unpaid work.

Generally, addressing the gender inequalities requires data that are timely and granular, in order for policies to take prompt and targeted actions. However, data provided by standard surveys (e.g. Labour Force Survey) on teleworking arrangements are often time-lagged and do not provide sufficiently detailed information at geographical level.

In fact, the recent literature has started using innovative data sources, such as mobile phone data, for analysing trends and providing predictions on various aspects of gender inequality, faster than the traditional data sources allow to. These analyses have shown that mobile phone data can be used to derive the gender and age distribution of population with relatively high accuracy \citep{Al-Zuabi2019,Jahanietal2017} and, therefore, to study mobility and spatial patterns by age and sex \citep{Caselli2020, Jo2020}. Other studies (\citep{JRC124130}) have used mobile phone data also to provide a timely and large scale picture on the effects of the mobility restrictions, imposed by the Italian authorities to fight the COVID-19 pandemic, on human mobility and their economic impact. Moreover, the literature has shown that the use of mobile phone data can be expanded to study and provide prediction regarding the employment levels \citep{Almaatouqetal2016, Sundosy2016, Toole2015}. 

This paper aims to contribute to this strand of data innovation literature by providing a method for nowcasting teleworking practices, disaggregated by sex and age, using mobile network operator (MNO) data. To do so, we first assess the relationship between mobility and teleworking; and in the second step, we nowcast the rate of teleworking by age and gender for 20 Italian regions. The case of Italy has been chosen only due to the availability of the data, nonetheless the methodology presented in this paper can be applied generally.

\section{Data}

The study relies on the combination of official statistics and innovative data sources. In specific, we use the Italian quarterly Labour Force Survey and the mobile network operator (MNO) data.

\subsection{LFS data} 

Data on teleworking and employment rates are based on the Italian cross-sectional quarterly Labour Force Survey (LFS), covering the period from Q1 to Q4 2020. The sample is restricted to the working-age population, aged 25-64, in order to exclude from the analysis students and retired individuals. The overall unweighted sample is composed of 187,000 observations for all four quarters of 2020.

Both employment rates and the share of teleworking were computed by sex, age and quarter for each of the twenty Italian regions. In addition to the traditional definition of employment rates, the variable ‘share of teleworking’ was defined as the share of people who declared to had been working from home for at least one day in the week prior to the interview, among the overall number of employed. 

The aggregated employment and teleworking data at the regional level were then merged with the mobility data.

\subsection{Mobility Data}
\label{sec:mob}

\begin{figure}[!htb]
\includegraphics[width=\textwidth]{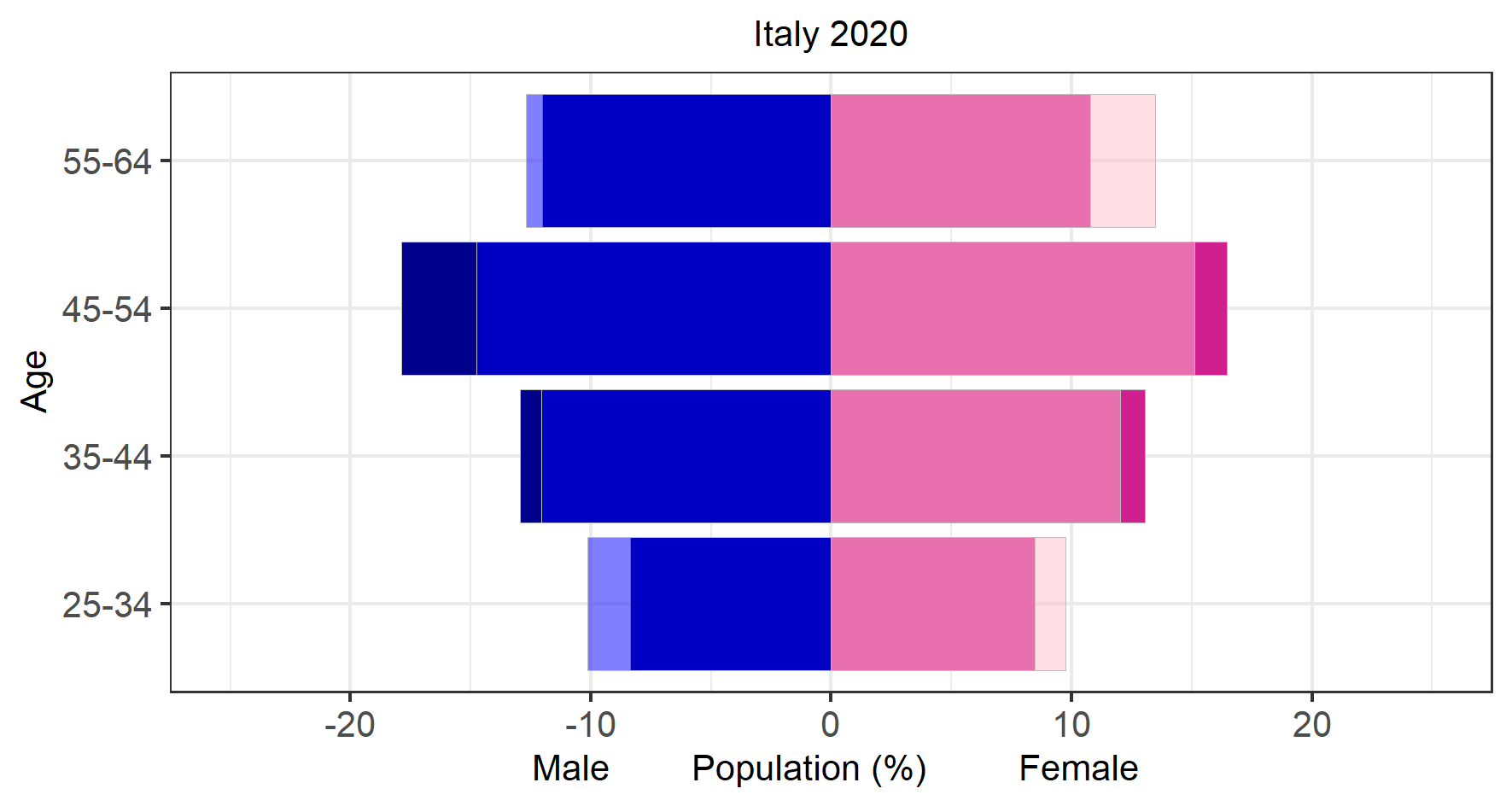}
\caption{Population pyramids for Italy in 2020. Population by age and sex according to Eurostat (Light-colored bars) and MNO subscribers by age and sex (dark-colored bars).}
\label{fig:demopyramid}    
\end{figure}

We used mobility data collected in the context of the Business-to-Government initiative between several mobile network operators in Europe and the European Commission \citep{vespe2021}. The purpose of this unprecedented collaboration is to make relevant information available to the European
Union, in order to predict and prevent the spread of COVID-19 and to manage related social,
political and financial impacts\footnote{See the Letter of Intent signed by the GSMA and the European Commission: \url{https://www.gsma.com/gsmaeurope/wp-content/uploads/2021/02/Letter-of-Intent_final_16-April-2021.pdf}.}.
The mobility data used in this study are in the form of aggregated and anonymized origin–destination matrices and are derived from mobile phones activity records. Some MNOs participating in the initiative provided origin–destination matrices disaggregated by demographic groups, while still preserving anonymity and non-identifiability of such groups. The Italian origin-destination matrices by age and gender have the following attributes:
\begin{itemize}
 \item origin municipality: prevalent area between 0:00 and 7:59;
 \item destination municipality: prevalent area between 8:00 and 15:59;
 \item 10-year age groups: 25-34, 35-44, 45-54, 55-64;
 \item gender: Male/Female;
 \item date: from 01 February 2020 to 31 August 2021;
 \item volume: number of mobile phone users who moved from the origin to the destination municipality. When the origin and the destination municipality is the same the volume describes users whose prevalent area remained the same.
\end{itemize}
Our sample is restricted only to consumer contracts and is derived from data generated on daily basis by almost 8 million mobile phone users, from which 52\% are male and 48\% are female. In the Figure \ref{fig:demopyramid} we compare the distribution of Italian population by age and sex according to Eurostat (light-colored bars) with the age and sex distribution of MNO subscribers having consumer contracts (dark-colored bars). The overlapped population pyramids suggest that the MNO data are largely representative of different age-sex population groups. An exception is that of men aged 45-54 who appear somewhat over-represented in the MNO data. 
Since the aim of this article is to identify changes in teleworking trends using the mobility trends, in the model we only use data on weekdays mobility (Monday to Friday). We assume that the ``\textit{place of residence}'' is the location where the subscribers spend most of their time in the first part of the day (00:00 - 07:59), and that the ``\textit{place of employment}'' is the location where the subscribers spend most of the second part of the day (8:00-15:59). This assumption, however, excludes from the sample people who are working in night-shifts or have non-typical working hours patterns. Therefore, our analyses are limited to the population with standard working hours patterns.

\begin{figure}[!htb]
 \includegraphics[width=\textwidth]{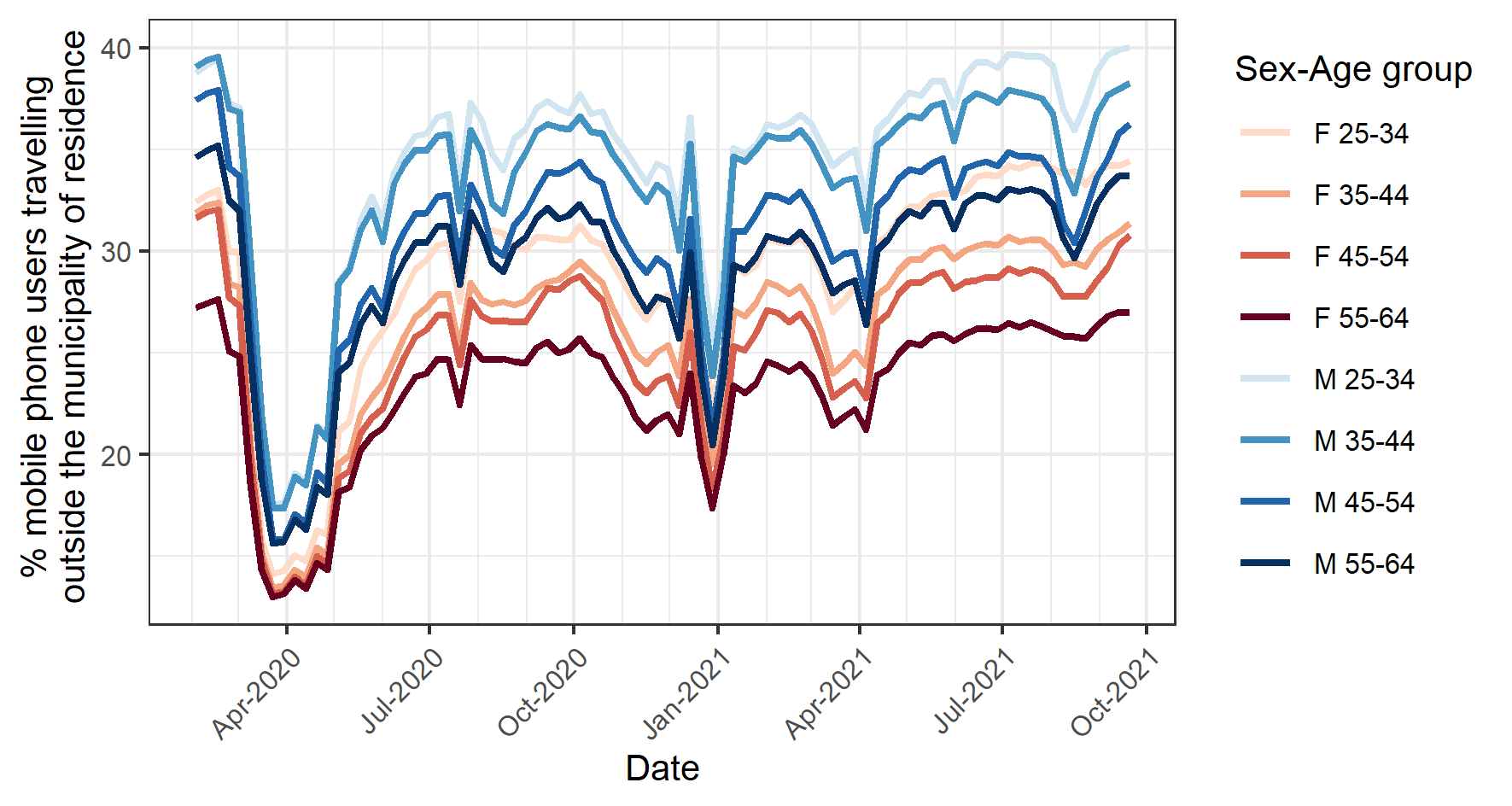}
\caption{Share of mobile phone users whose predominant location between 8:00 to 15:59 during the weekdays was outside the municipality of their residence}
\label{fig:mobilityagesex}    
\end{figure}
Moreover, it should be stressed out that our data allow capturing only the inter-municipal mobility, that is the mobility from one municipality/commune to another (or sub-municipality regions in the case of Milan and Rome). Movements that occur within the same municipality, e.g. when a person lives and works in the same municipality are not registered in our data, and therefore it is not possible to distinguish short distance mobility from immobility. According to \cite{ISTAT2019}, higher shares of women work in the same municipality in comparison to men, although the share considerably varies across regions.
In Figure \ref{fig:mobilityagesex} we present the share of mobile phone users that spend most of the second part of the weekdays (Monday to Friday, 8:00 to 15:59) outside the municipality of their residence. Figure \ref{fig:mobilityagesex} shows that, in terms of inter-municipal mobility, men are more mobile than women across all age groups and that the mobility of both sexes is returning to the pre-pandemic levels.

\newpage
\section{Methodology}

The method is based on a linear-log regression model which estimates the share of Italians teleworking by age, sex, region and quarter, as a function of the employment rate, the log of mobility rate and the region-sex fixed-effects. We estimate, therefore, the following equation \ref{eq:mod1}.
\begin{equation}
teleworking_{s,a,r,q} = \alpha + \beta\cdot {\tt region}_r : {\tt Sex}_{s} + \gamma \cdot \log{\tt mobility}_{s,a,r,q} + \delta \cdot {\tt employment}_{s,a,r,q} + \epsilon_{s,a,r,q}
\label{eq:mod1}
\end{equation}
where:
\emph{teleworking}: is the share of people who teleworked at least one day in the week prior to the interview in the total number of people employed. This share is estimated for each age group, sex, region and quarter;
\emph{mobility}: is the share of people moving in the total number of subscribers. It is estimated as the mean of daily share of mobility by age group, sex, region and quarter;
\emph{employment}: is the share of people employed in the total population and it is available for each age group, sex, region and quarter;
\emph{Sex} and \emph{s} is the sex [Male; Female];
\emph{Region} and \emph{r} is the Italian NUTS2 Region at [Lombardy, Piemonte...etc];
\emph{a} is the age group [25-34;35-44,45-54,55-64];
\emph{q} is the quarter [Q3 2020, Q4 2020];
$\epsilon_{s,a,r,q}$ is the sequence of independent and identically distributed Gaussian random variables.

In the model, we use logged mobility due to a non-linear relationship between mobility and teleworking. This non-linear relationship is conditioned by the model's definition of ``\textit{teleworking}'' as having worked from home for at least one day per week. Our assumption is that higher teleworking rates reflect teleworking practices beyond one day per week, resulting proportionally in a higher degree of mobility reduction. We test this assumption by comparing the model in equation \ref{eq:mod1} with the same model where mobility is not logged. The non-linear relationship was confirmed since the log mobility increased the coefficient of determination  of the model from $R^{2}$=0.589 to $R^{2}$=0.606.

Moreover, we include the region-sex fixed-effects in order to capture the gendered-regional differences in the percentage of employees working outside the municipality of residence. These gendered-regional differences in mobility have been previously captured by the Italian permanent census of population \cite{ISTAT2019}.

We run the regression using data only for the period Q3 and Q4 of 2020. The Q1 2020 was not included in the model because the mobility data does not cover the entire quarter, but only the last two months (01 February to 31 March 2020). The Q2 2020 was also excluded from this study as it covers the lockdown period when most of the businesses were closed and the inter-municipal movements were banned. 

Finally using the model in equation \ref{eq:mod1} and the coefficients presented in Table \ref{minimal model} we predicted the teleworking rates by age, sex and each region of Italy for the quarters Q1-Q3 of 2021 where official data about teleworking are not yet available. The nowcasted values are shown as dotted red and blue lines in Figures \ref{fig:Lombardia-tele}, \ref{fig:Toscana-tele} and \ref{fig:Calabria-tele}. Since we had no updated employment rate data for the quarter Q1, Q2 and Q3 of 2021 we kept employment rate fixed to the one of Q4 of 2020. Other scenarios on employment change can be applied equally.

\section{Results} 

Italy represents a unique case study. First, prior to the COVID-19 pandemics, Italy was the country with one of the lowest teleworking rates in the EU (below 5\% in comparison teleworking rates higher than 30\% Sweden, the Netherlands, Luxembourg and Finland \citep{Milasi2021}). This difference between Italy and the rest of the EU countries has been attributed to between-sector differences – i.e. to the different share of employment in occupations more amenable to remote work \citep{DINGEL2020} –, but even more so to within-sector differences – to i.e. differences in the organisational, managerial and business cultures among regions and countries \citep{Milasi2021}. Secondly, Italy was the first European country to be hit by COVID-19 pandemic and also the first EU country to implement a national lockdown on 9 March 2020. As a result, these circumstances lead to a rapid shift toward teleworking arrangements in a country with a strong tradition of office working.

\subsection{Assessing the relationship between mobility and teleworking}

In Table \ref{minimal model} we present the results of the model. First, the model shows a negative association between mobility and teleworking: i.e. the higher the mobility rate is, the lower is the level of teleworking. Secondly, given a fixed mobility, an increase in the occupation rate is expected to have a positive impact on the level of teleworking.

\input{model_minimal}

\begin{figure}[!htb]
 \includegraphics[width=1\textwidth]{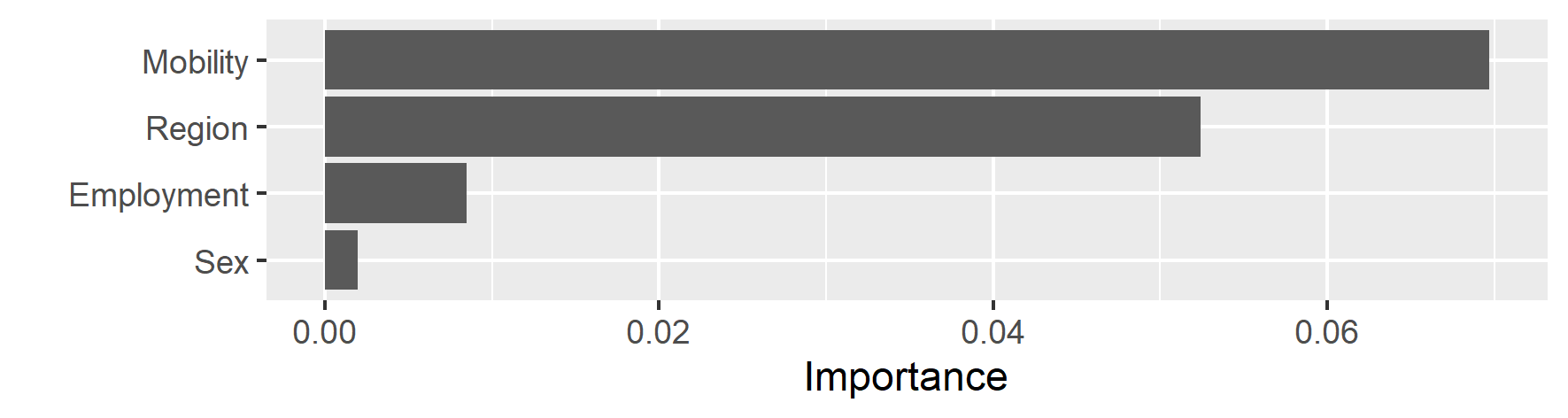}
\caption{Importance of model's variables}
\label{fig:VariableImportance}    
\end{figure}

In Figure \ref{fig:VariableImportance} we present the variable importance estimate which shows that mobility plays the most important role. Sex plays a minor role in the model since the mobility data are disaggregated by sex and age group, and thus are capturing well that females are moving less than males and consequently teleworking to a higher degree than the men do.

\begin{figure*}[!h]
 \includegraphics[width=1\textwidth]{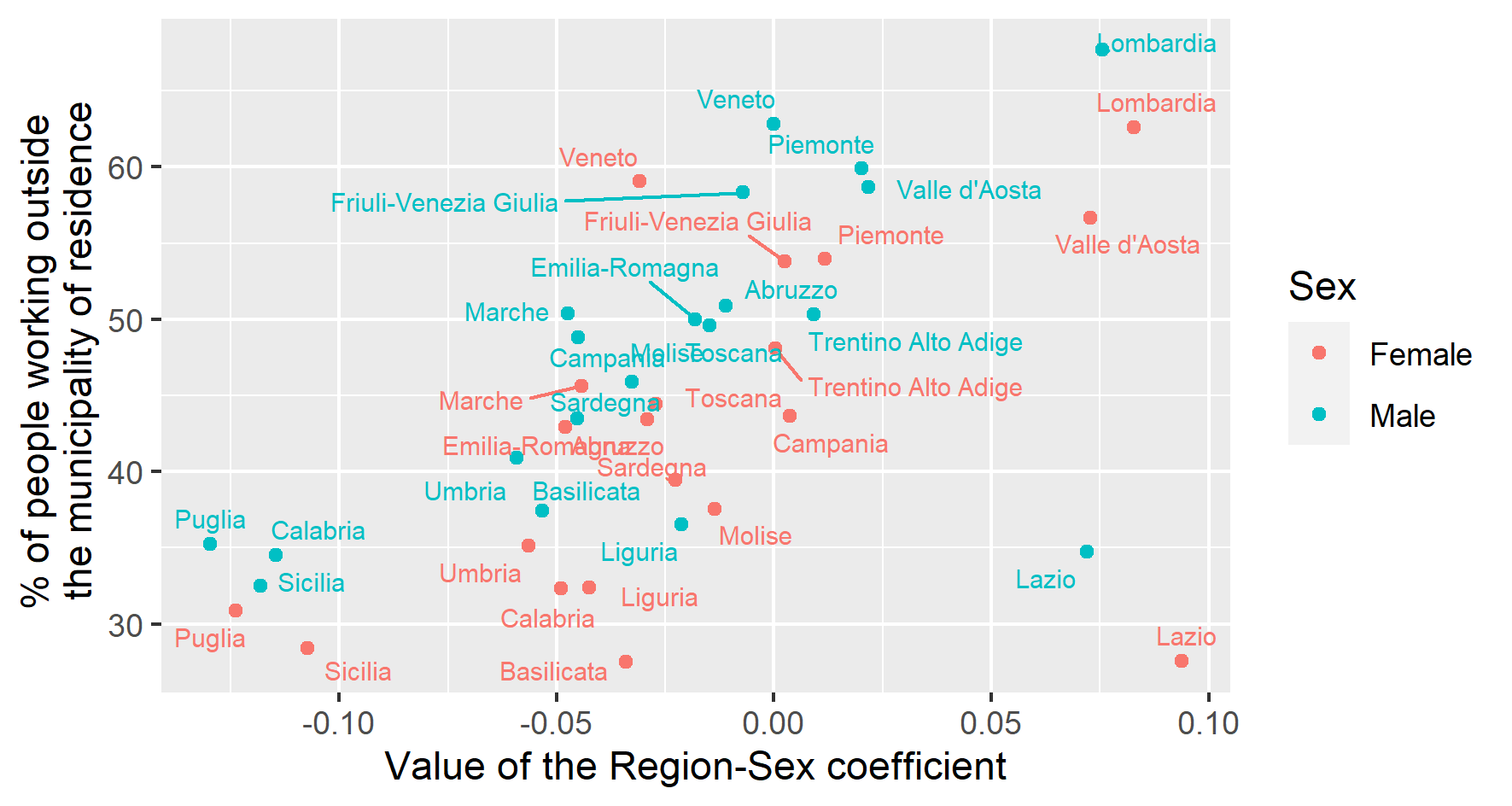}
\caption{Coefficient of the model Vs \% of worker that work outside the municipality of their residence}
\label{fig:coefficients}    
\end{figure*}

\begin{figure*}[!h]
\includegraphics[width=1\textwidth]{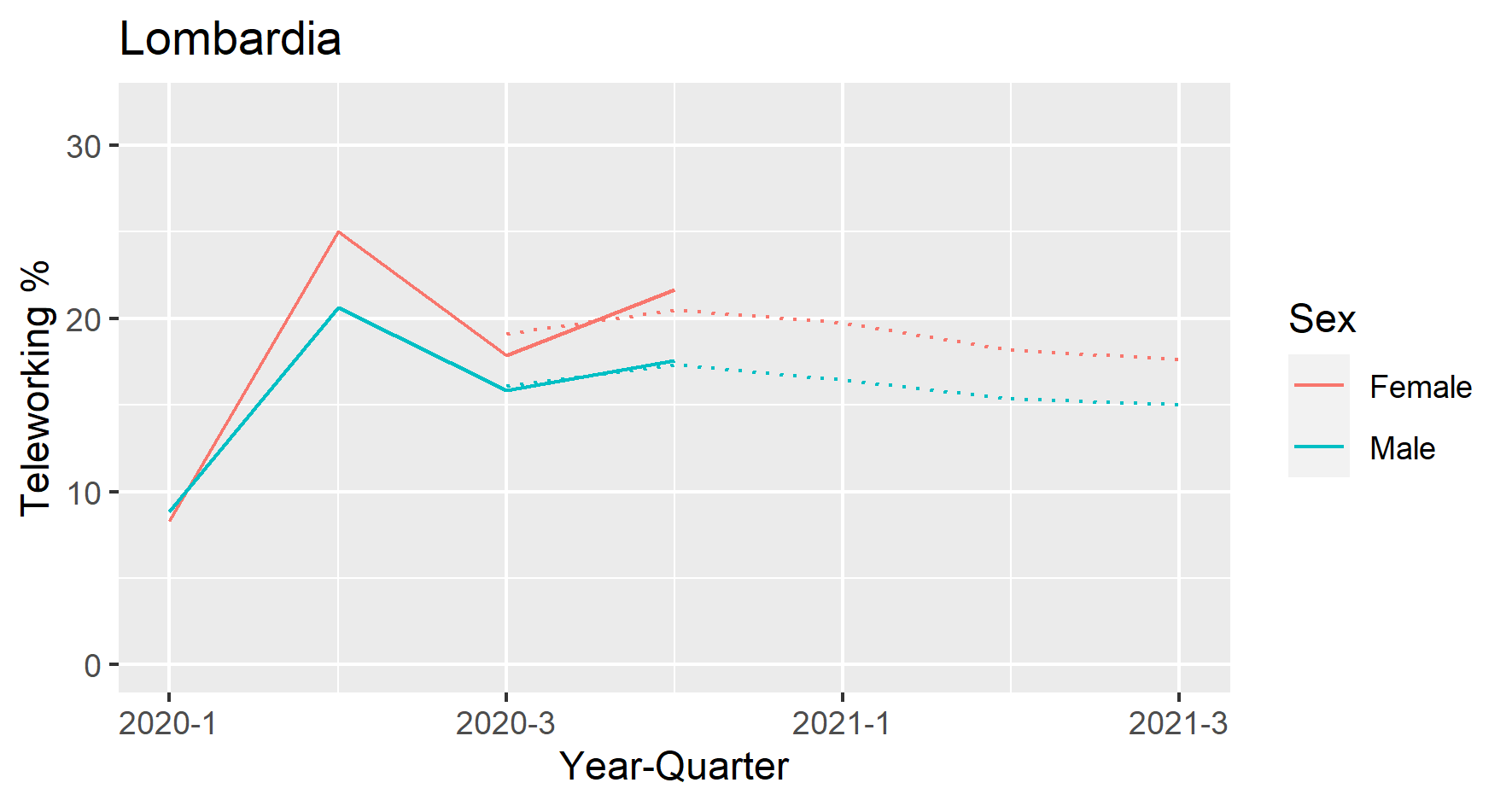}
\caption{Teleworking rate in Lombardia, Labour Force Survey teleworking estimates in solid lines, mobility-derived teleworking estimates in dotted lines}
\label{fig:Lombardia-tele}    
\end{figure*}

\begin{figure*}[!h]
\includegraphics[width=1\textwidth]{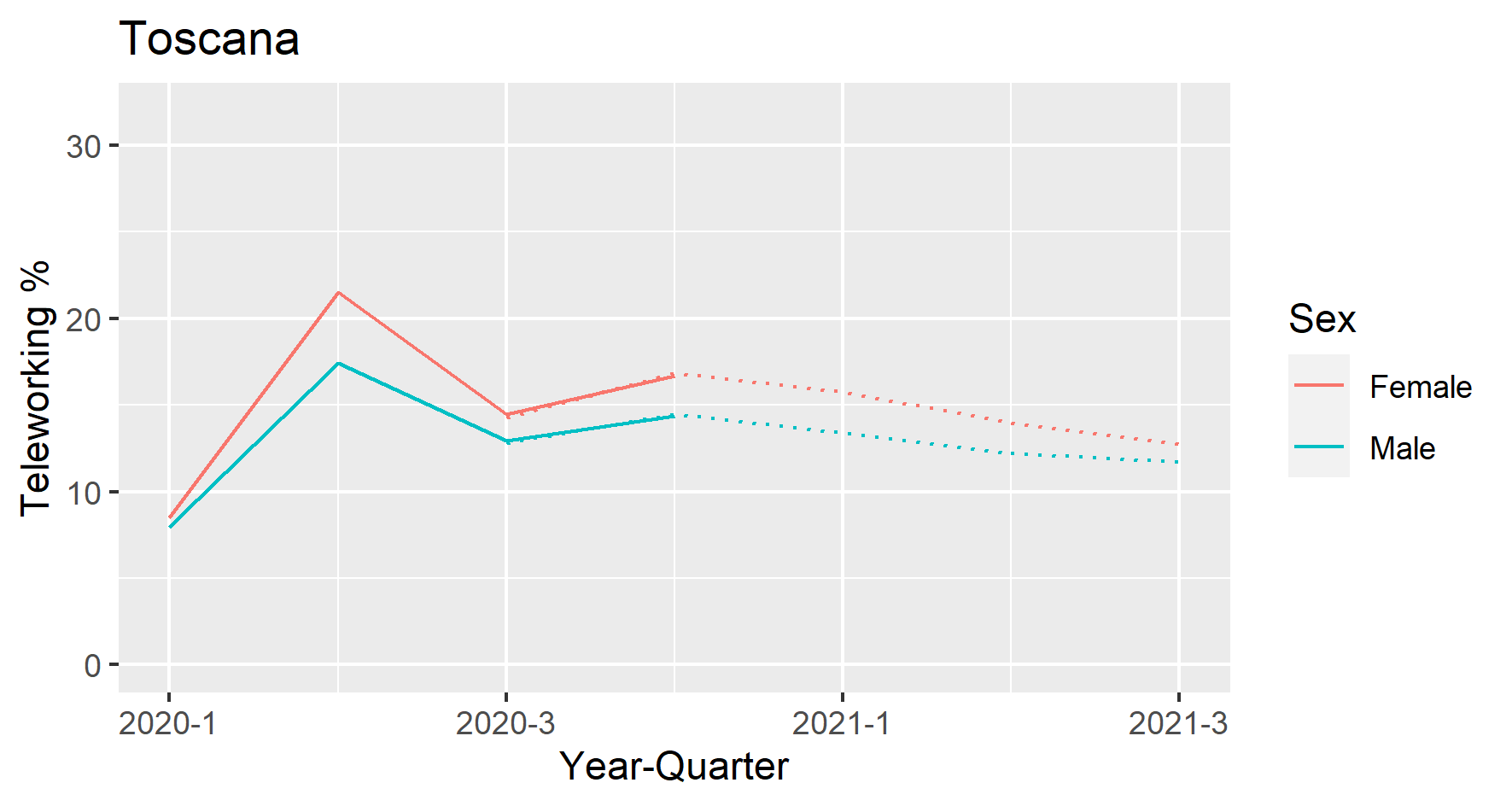}
\caption{Teleworking rate in Toscana, Labour Force Survey teleworking estimates in solid lines, mobility-derived teleworking estimates in dotted lines}
\label{fig:Toscana-tele}    
\end{figure*}

\begin{figure*}[!h]
\includegraphics[width=1\textwidth]{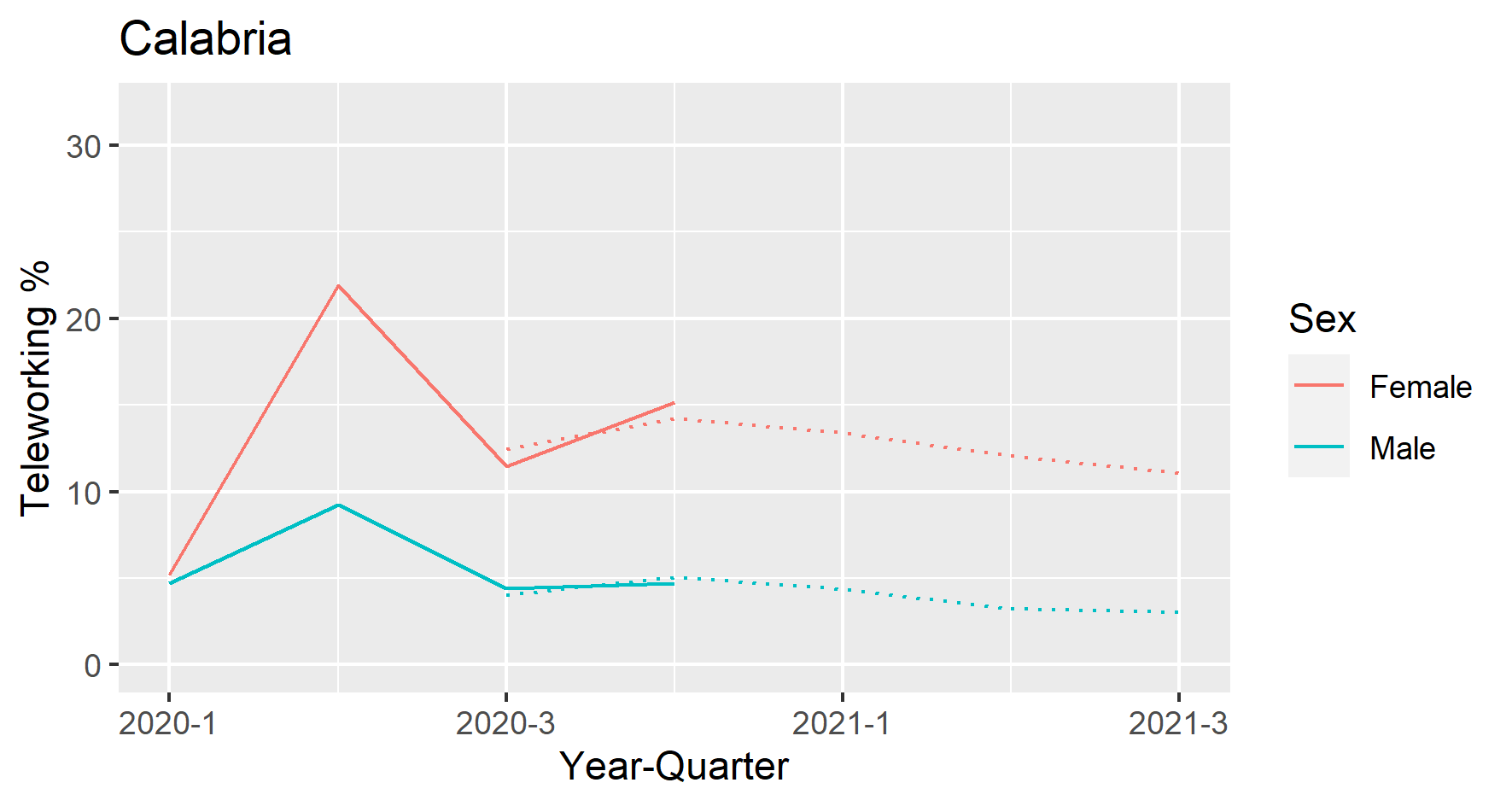}
\caption{Teleworking rate in Calabria, Labour Force Survey teleworking estimates in solid lines, mobility-derived teleworking estimates in dotted lines}
\label{fig:Calabria-tele}    
\end{figure*}

As shown in Figure \ref{fig:coefficients} the dummy variables for each region and sex are correlated well with the percentage of men and women working outside their municipality of residence in 2019, where the latter is captured by data on population work-related mobility patterns drawn from the Italian permanent census of population \cite{ISTAT2019}. Lombardia and Lazio are special cases since in Milan and Rome movements within the municipalities are also captured by MNO data, as long as the subscriber is moving from one sub-municipality region to another (9 sub-municipality regions in Milan and 108 sub-municipality regions Rome municipality). This explains also why the dummy variables for Lazio and Lombardia regions, where Rome and Milan are located respectively, have very high coefficient values. If we remove the regions of Lombardia and Lazio then the Pearson correlation between the dummies variables and the share of workers moving to another municipality for work as recorded by \cite{ISTAT2019} is $\rho$ = 0.72 while the Spearman correlation is $\rho$ = 0.74. Such correlations prove the robustness of our model and that the region-sex dummy variables describe well the employment reality, as this is also reflected in the official statistics. Finally, as shown in Figures \ref{fig:Lombardia-tele}, \ref{fig:Toscana-tele} and \ref{fig:Calabria-tele} the proposed model captures well the actual variation in teleworking over the Q3 and Q4 of 2020. This robustness of our model is confirmed also for the remaining regions (see Annex).

\subsection{Explaining patterns \& teleworking trends   by sex, age and region}

\paragraph{Actual teleworking trends in Q1-Q4 of 2020}\mbox{}

The Italian LFS data gives insights on trends in actual teleworking rates for the Q1-Q4 of 2020 (full red and blue lines in Figures \ref{fig:Lombardia-tele}, \ref{fig:Toscana-tele} and \ref{fig:Calabria-tele}). The starting point of our analysis is the Q1 of 2020 which covers months (January-February) prior to the COVID-19 crisis in Italy and the month (March) of the first national lockdown. In Q1 2020 we observe that all regions registered relatively low teleworking rates, between 5\%-10\%, and with fairly similar patterns between genders within regions. The only exception are three Southern regions - Campania, Basilicata, and Sicilia – where the female teleworking was much more expanded in comparison to the male one. 
The Q2 2020 registered a teleworking boom for both genders, and more so among women, leading to a considerable teleworking gender gap in all Italian regions. For example, in Lazio the share of women teleworking went from 8\% in Q1 2020 to 30\% in Q2 2020 while the share of men teleworking increased from 8\% in Q1 2020 to 22\% in Q2 2020. This teleworking boom, however, appears limited to the Q2 2020 as the following quarters Q3-Q4 register substantial decrease in teleworking rates.
In Q4 2020 – the last quarter in which we observe the actual teleworking shares – generally the share of teleworking men and women appears higher than in Q1 2020, yet substantially lower than during the teleworking boom of Q2 2020. Moreover, in certain regions (Abruzzo, Calabria, Marche, Molise, and Umbria) the share of men teleworking returned to the values registered in Q1 2020. This suggests that, according to latest official statistics, men started returned physically to work more than women.

\paragraph{Nowcasted teleworking trends in Q1-Q3 of 2021}\mbox{} 

\begin{figure*}[!h]
\includegraphics[width=1\textwidth]{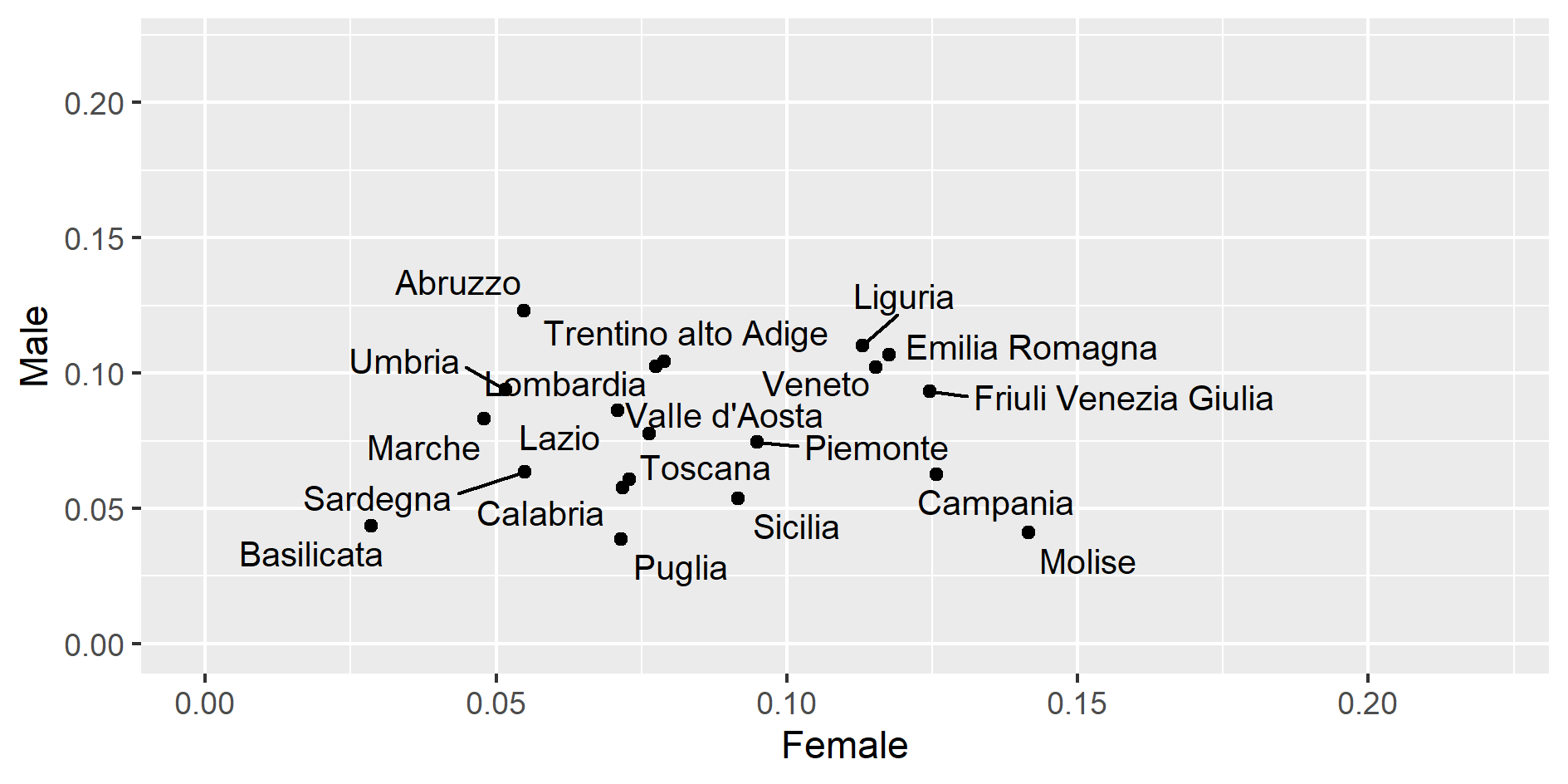}
\caption{teleworking rate by sex of workers age 35-44 for the Q1-2020}
\label{fig:teleworkingQ12020}    %
\end{figure*}

\begin{figure*}[!h]
\includegraphics[width=1\textwidth]{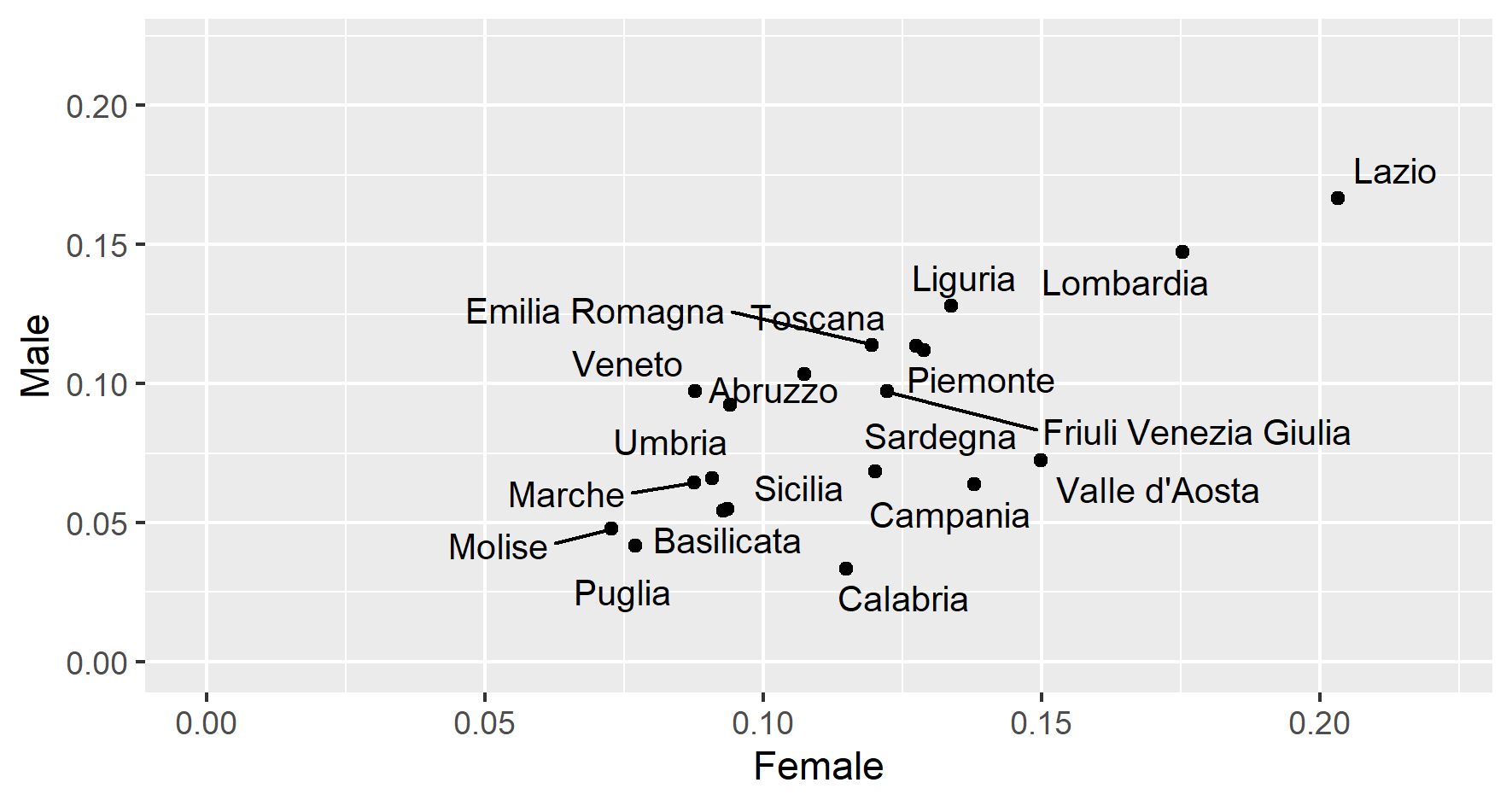}
\caption{teleworking rate by sex of workers age 35-44 for the Q3-2021}
\label{fig:teleworkingQ32021}    %
\end{figure*}

According to our nowcasted results for Q1-Q3 of 2021 (dotted red and blue lines in Figures \ref{fig:Lombardia-tele}, \ref{fig:Toscana-tele} and \ref{fig:Calabria-tele}). the share of teleworking among men and women slightly reduced in comparison to Q4 2020. 
When comparing Q1 2020 in Figure \ref{fig:teleworkingQ12020} with nowcasted values for Q3 2021 in Figure \ref{fig:teleworkingQ32021}, we observe that the share of men teleworking in Q3 2021 returned to the values of Q1 2020 in 13 out of 20 regions \footnote{In case of men these include Marche, Molise, Abruzzo, Umbria, Calabria, Veneto, Puglia, Sardegna, Emilia Romagna, Liguria, Sicilia, Trentino Alto-Adige and Campania}; whereas, the share of women teleworking in Q3 2021 returned to the level of Q1 2020 only in 4 regions \footnote{In case of women these include Abruzzo, Basilicata, Veneto and Emilia Romagna}. These results seem to confirm the trend of men returning to work at a higher pace than women.

Looking at the extent of teleworking gender differences in Q3 2021, we can broadly distinguish regions in three clusters. 
On one side, there is the cluster of regions with relatively high teleworking gender gap: around 7 p.p. in Calabria, Campania and Valle D’Aosta and around 5 p.p. in Puglia, Sardegna, Sicilia and Basilicata. It should be noted that, unlike Campania, Sicilia and Basilicata which had gendered-teleworking practices also prior to COVID-19, in the remaining regions the teleworking gender gap has been widened since the teleworking boom in Q2 2020. On the opposite side, there is the cluster of regions that have reached or are very close to reaching (gender gap lower than 1\%) the teleworking gender parity: Abruzzo, Emilia Romagna, Liguria, Toscana, Trentino Alto Adige and Veneto. The cluster of remaining regions can be categorised as a cluster of low teleworking gender gap (between 1-4 p.p.) and, observing the trend, it can be assumed that these regions will close the teleworking gap in near future and in absence of new shocks. 
Moreover, we look at differences in teleworking gender gaps among four age groups. Generally, for both women and men, it can be observed that as the age increases, the share of workers teleworking increases as well. In fact, we find the lowest teleworking share among the age group 25-34 in all regions; whereas the highest teleworking share is generally registered among 55-64, although some exceptions exist where the rate appears the highest for the age group 45-54. The COVID-19 crisis appears to have affected similarly all age groups and as a result, the widening of teleworking gender gaps has occurred across all age groups.

\section{Discussion and Conclusions}
\label{sec:discussionconclusions}
At the eve of the COVID-19 pandemic, Italy was a country with a strong tradition of office working and low, generally gender-equal, teleworking patterns. The results of our analysis confirm that the COVID-19 lead to a teleworking boom in Q2 2020 as well as to a widening of teleworking gender gaps, as higher shares of women have adopted teleworking in comparison to men. 
The teleworking boom, however, appears to be limited to Q2 2020. In fact, the trends in teleworking rates – emerging from both actual and nowcasted values - indicate that since Q2 2020 the share of workers teleworking has reduced. According to our nowcasted values, in Q3 2021 the share of men teleworking returned to the values of Q1 2020 in 13 out of 20 regions. At the same time, the share of women teleworking – despite its decline since the teleworking boom quarter – reached the level of Q1 2020 only in 4 regions. Overall, these results suggest that, generally, in Italy men have been returning to office work at a higher pace than women. In addition, in some Southern regions the teleworking gender gap, generated in most cases by the teleworking boom, still remains relatively large. This finding calls for additional analyses in order to understand the underlying determinants and circumvent the risk that, as \citet{rubery2020covid} warn: ``\textit{New forms of gender segregation could emerge if women are not only expected to telework but in fact remain home-based workers while men return to the office}''.
Our findings, also contribute to the broader discussion on the uptake of teleworking. The COVID-19 pandemic has proven the feasibility of expanding teleworking practices in all Italian regions as well as the potential advantages associated with it, such as firm savings and its reinvestments in innovation. Nonetheless, a wider adoption of teleworking practices in Italy did not occur beyond year 2020. In order to reap the benefits of teleworking, these findings indicate a need for additional analyses to better understand the barriers limiting the uptake of teleworking arrangement, and the extent to which these barriers are linked to between- and within-sector differences.
Finally, in terms of methodology, this paper has shown the potential of using mobile network operators’ data also for gaining almost real-time insights on the trends and patterns of flexible working arrangements, such as teleworking, which can feed in a timely manner policy interventions.

\section*{Acknowledgments}
The authors acknowledge the support of European MNOs (among which 3 Group - part of CK Hutchison, A1 Telekom Austria Group, Altice Portugal, Deutsche Telekom, Orange, Proximus, TIM Telecom Italia, Tele2, Telefonica, Telenor, Telia Company and Vodafone) in providing access to aggregate and anonymised data, an invaluable contribution to the initiative. The authors would also like to acknowledge the GSMA\footnote{GSMA is the GSM Association of Mobile Network Operators.}, colleagues from Eurostat\footnote{Eurostat is the Statistical Office of the European Union.} and ECDC for their input in drafting the data request. 
Finally, the authors would also like to acknowledge the support from JRC colleagues, and in particular the E3 Unit, for setting up a secure environment to host and process of the data provided by MNOs, as well as the E6 Unit (the ``\textit{Dynamic Data Hub team}'') for their valuable support in setting up the data lake.

\section*{Declarations}

\begin{itemize}
\item Funding: The authors received no specific funding for this work.
\item Conflict of interest: On behalf of all authors, the corresponding author states that there is no conflict of interest.

\end{itemize}

\bibliography{refbib}
\bibliographystyle{apalike}

\begin{appendices}
\section*{Teleworking Rates by Region}

\begin{figure*}[!h]
\includegraphics[width=1\textwidth]{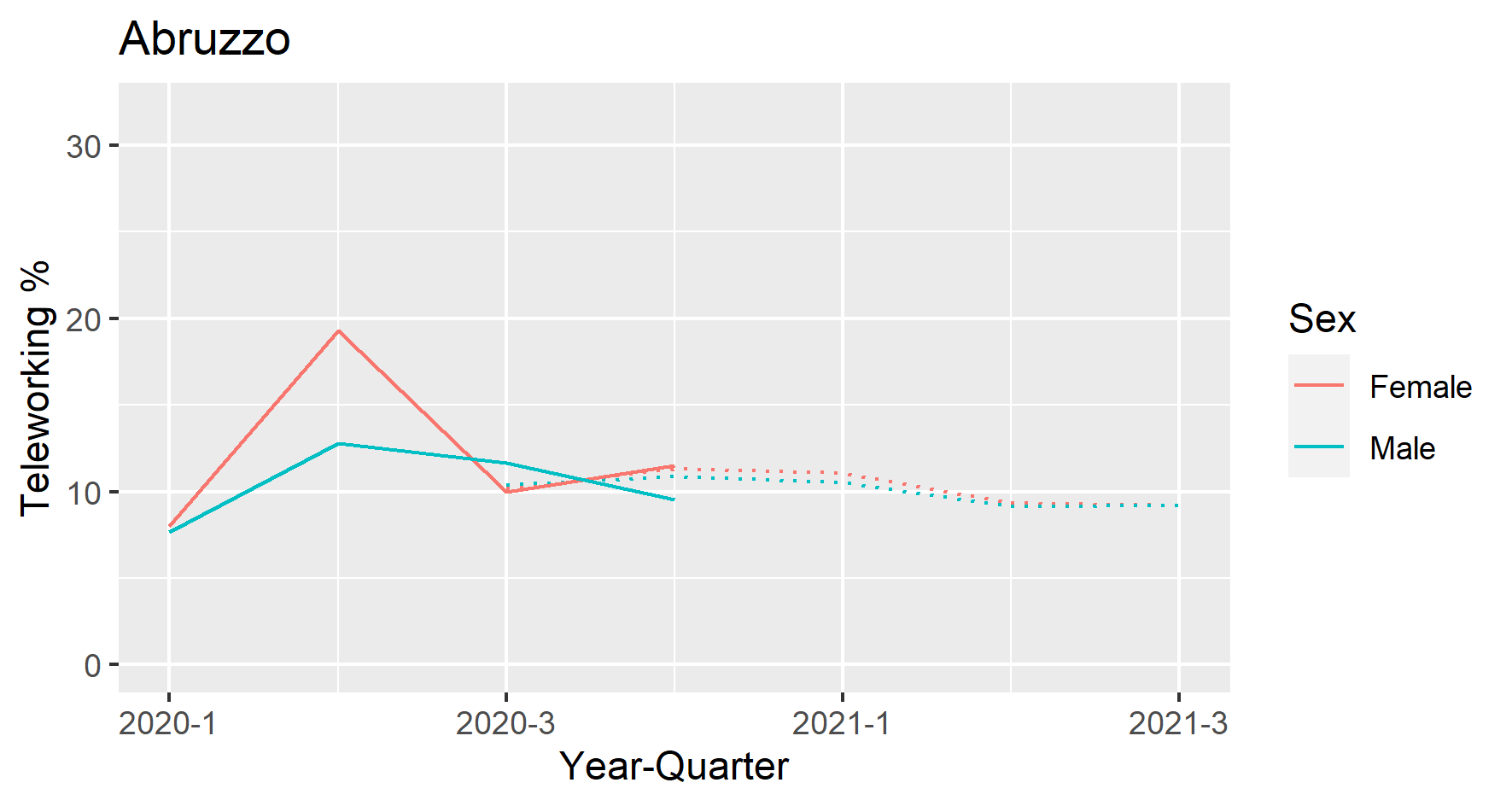}
\caption{Teleworking rate in Abruzzo,  Labour Force Survey teleworking estimates in solid lines, mobility-derived teleworking estimates in dotted lines}
\end{figure*}

\begin{figure*}[!h]
\includegraphics[width=1\textwidth]{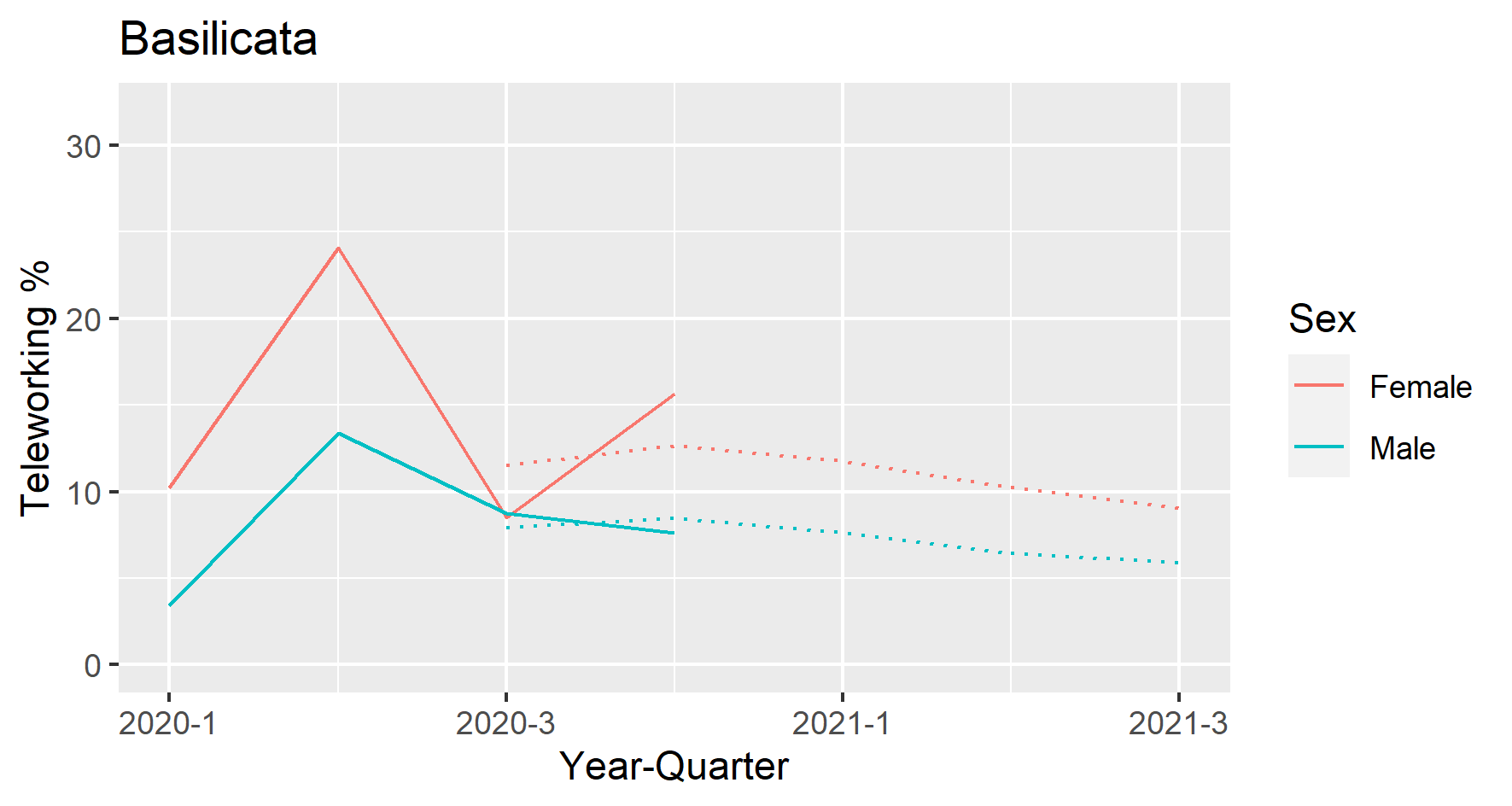}
\caption{Teleworking rate in Besilicata,  Labour Force Survey teleworking estimates in solid lines, mobility-derived teleworking estimates in dotted lines}
\end{figure*}

\begin{figure*}[!h]
\includegraphics[width=1\textwidth]{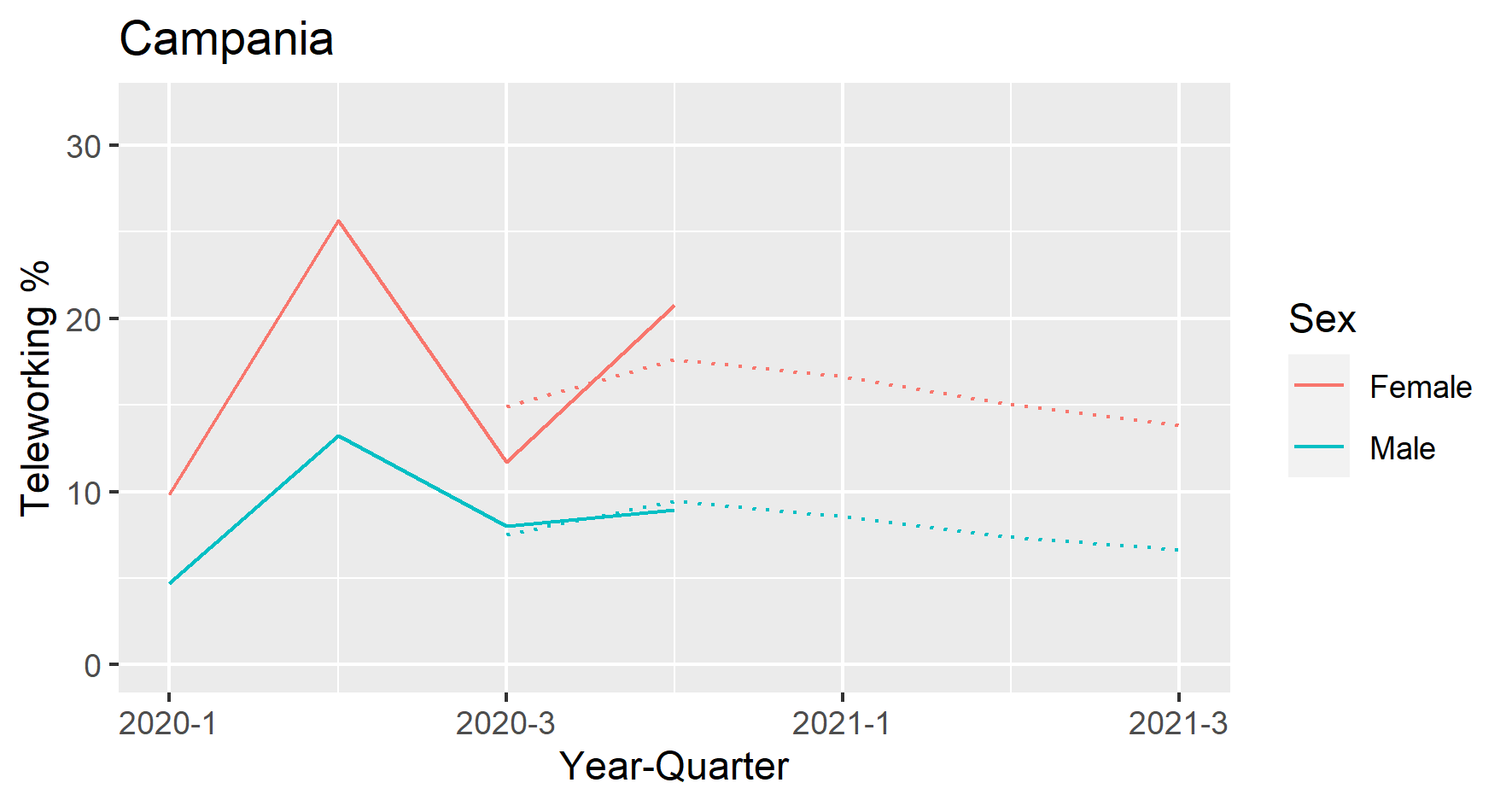}
\caption{Teleworking rate in Campania,  Labour Force Survey teleworking estimates in solid lines, mobility-derived teleworking estimates in dotted lines}
\end{figure*}

\begin{figure*}[!h]
\includegraphics[width=1\textwidth]{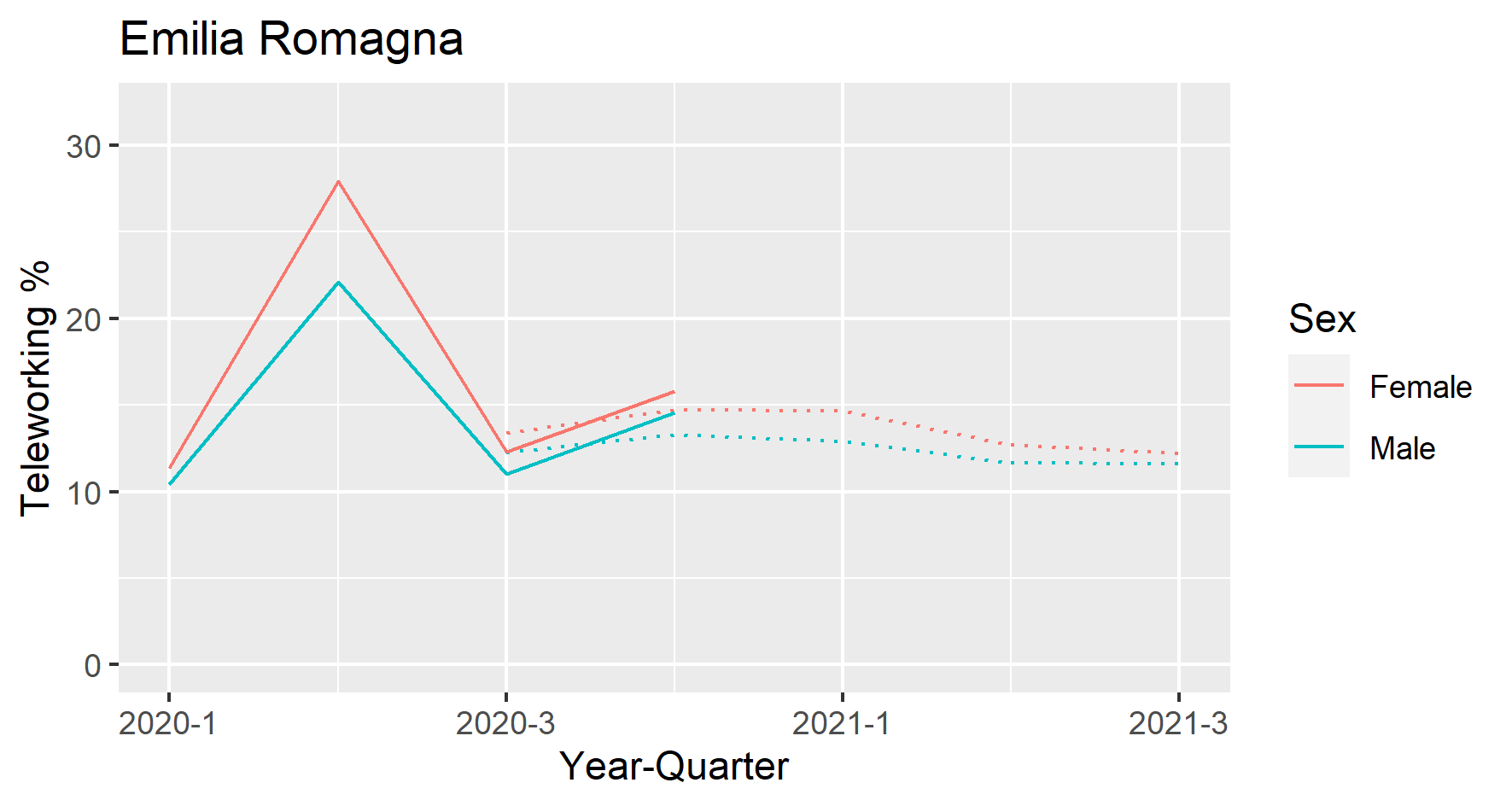}
\caption{Teleworking rate in Emilia Romagna,  Labour Force Survey teleworking estimates in solid lines, mobility-derived teleworking estimates in dotted lines}
\end{figure*}

\begin{figure*}[!h]
\includegraphics[width=1\textwidth]{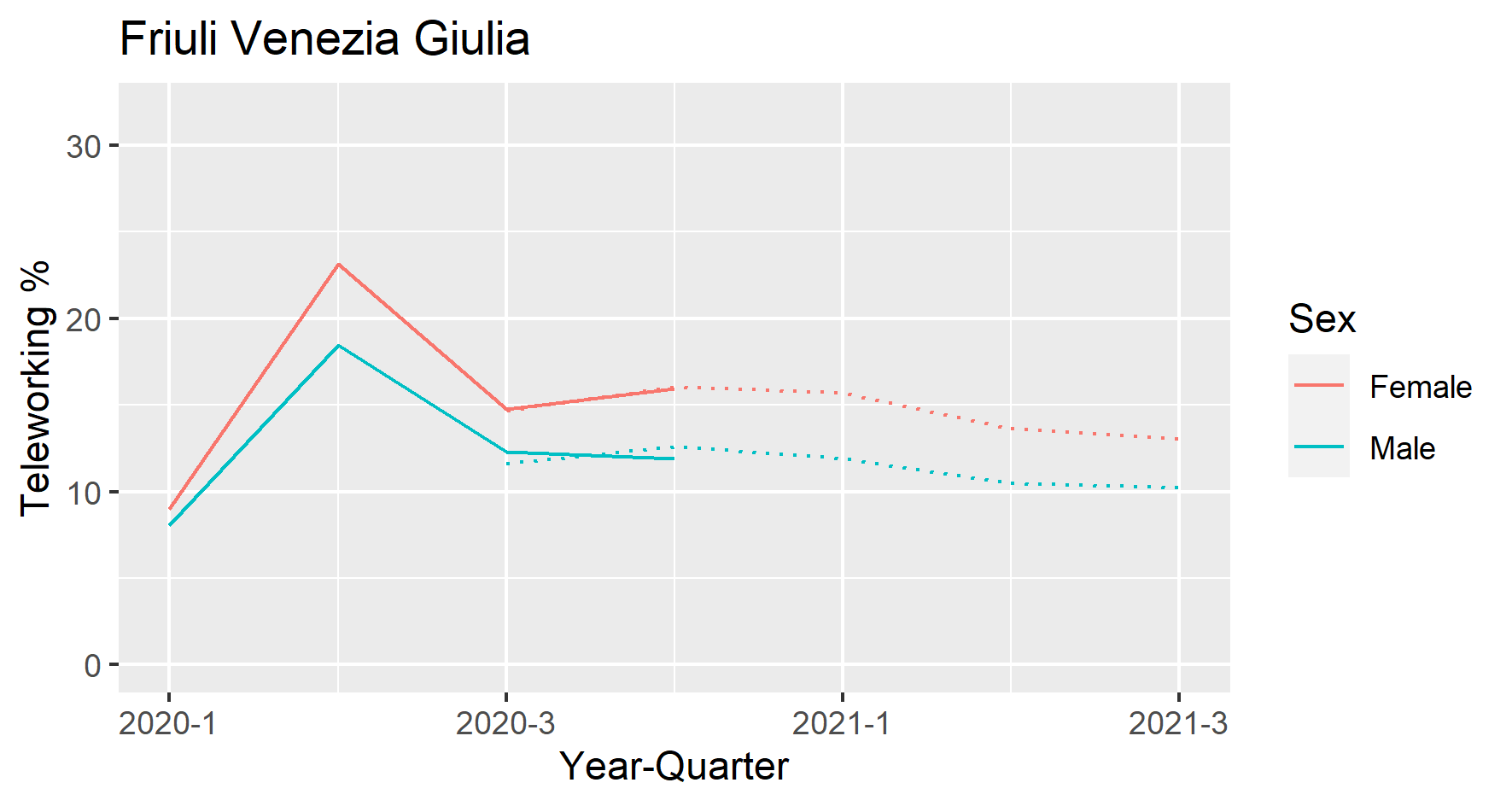}
\caption{Teleworking rate in Friuli Venezia Giulia,  Labour Force Survey teleworking estimates in solid lines, mobility-derived teleworking estimates in dotted lines}
\end{figure*}

\begin{figure*}[!h]
\includegraphics[width=1\textwidth]{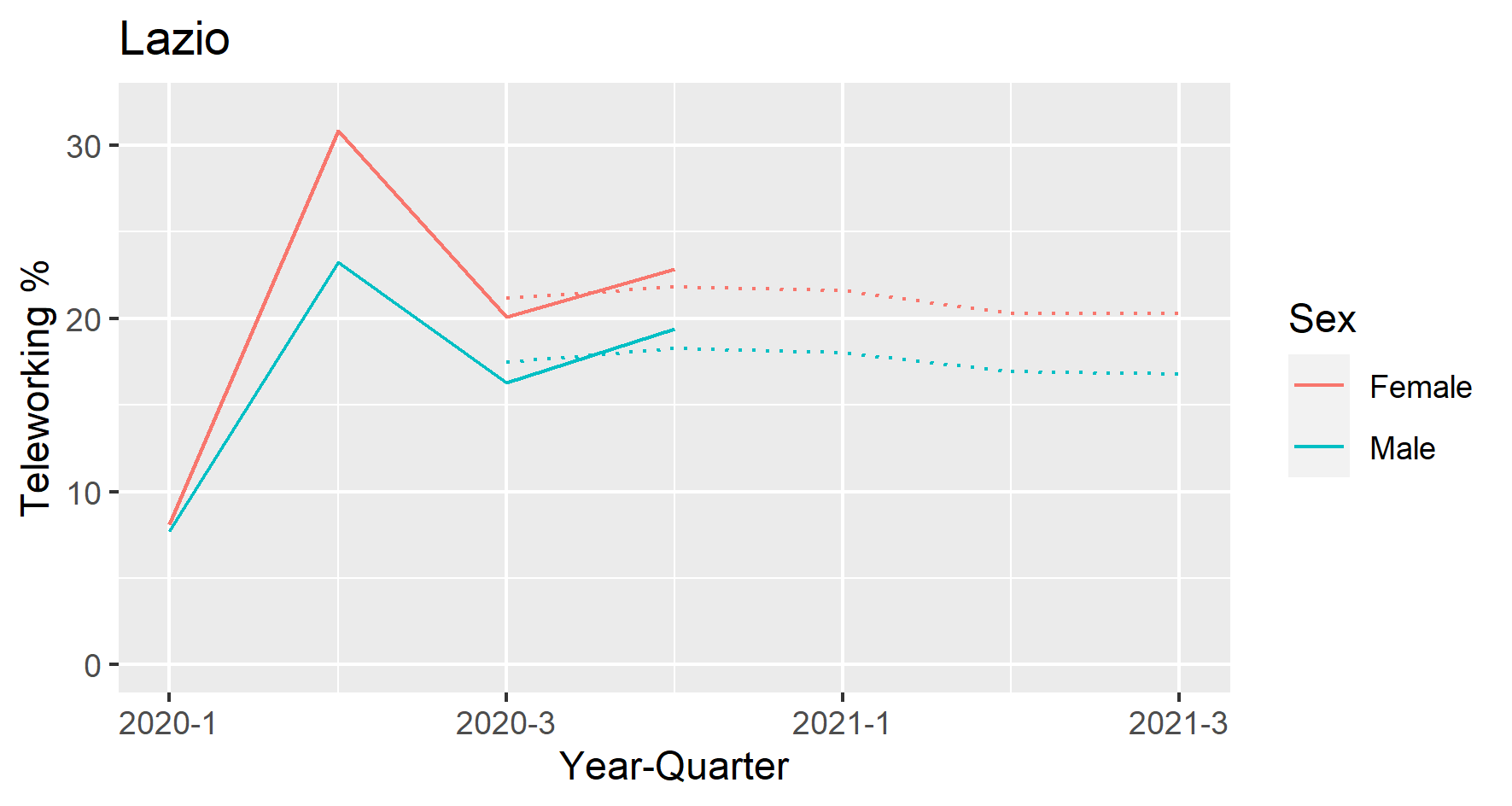}
\caption{Teleworking rate in Lazio,  Labour Force Survey teleworking estimates in solid lines, mobility-derived teleworking estimates in dotted lines}
\end{figure*}

\begin{figure*}[!h]
\includegraphics[width=1\textwidth]{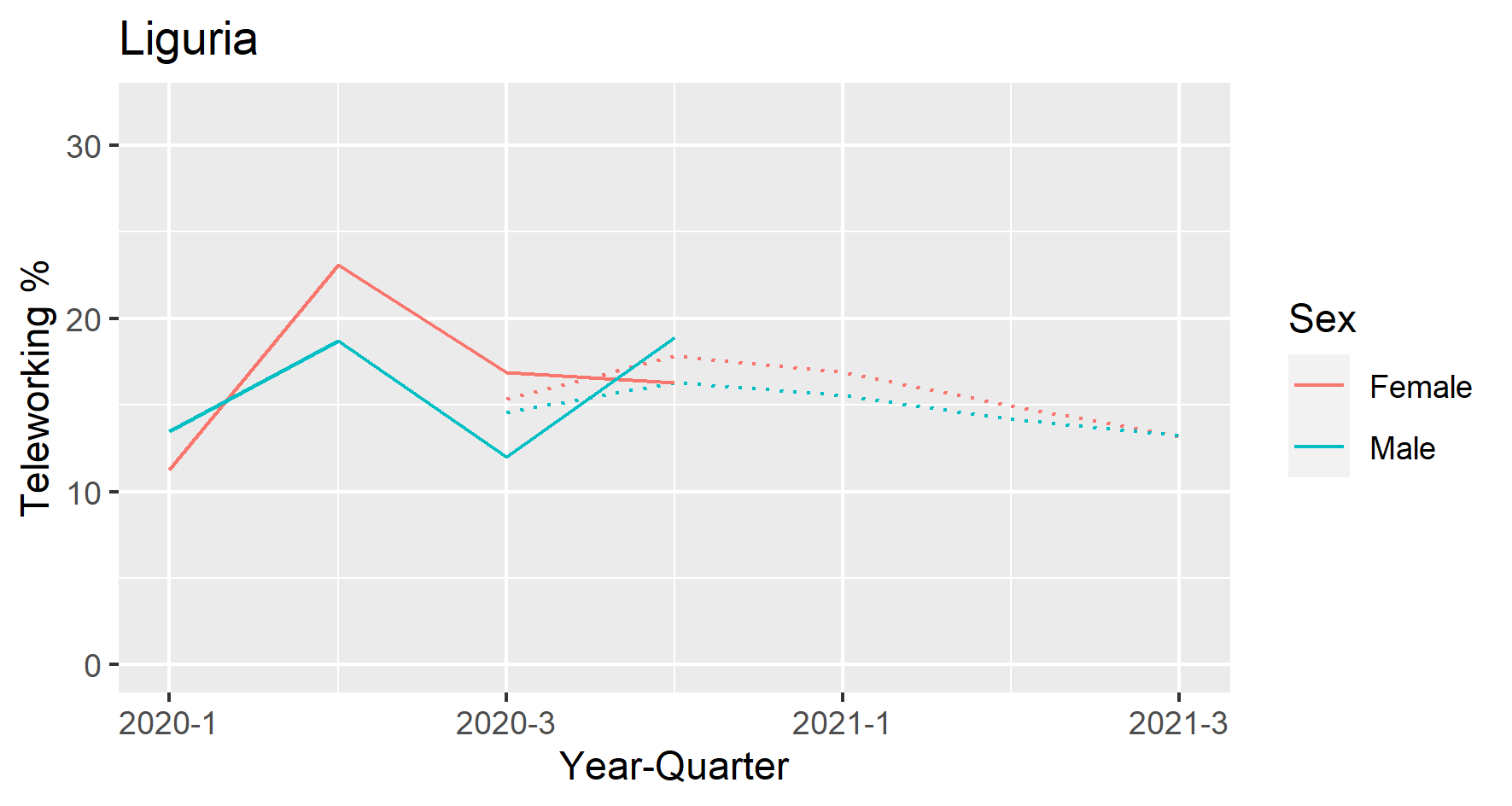}
\caption{Teleworking rate in Liguria,  Labour Force Survey teleworking estimates in solid lines, mobility-derived teleworking estimates in dotted lines}
\end{figure*}

\begin{figure*}[!h]
\includegraphics[width=1\textwidth]{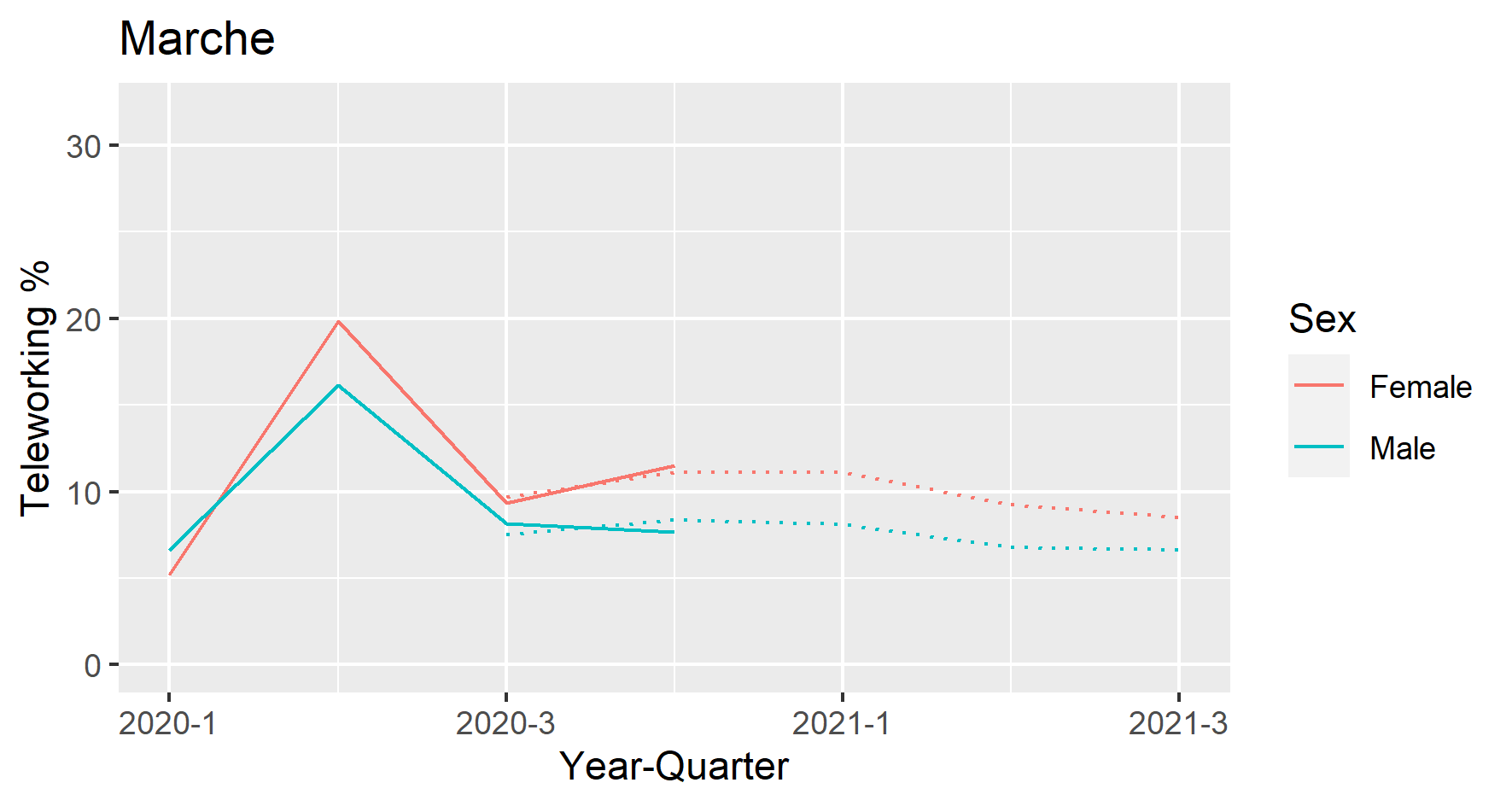}
\caption{Teleworking rate in Marche,  Labour Force Survey teleworking estimates in solid lines, mobility-derived teleworking estimates in dotted lines}
\end{figure*}

\begin{figure*}[!h]
\includegraphics[width=1\textwidth]{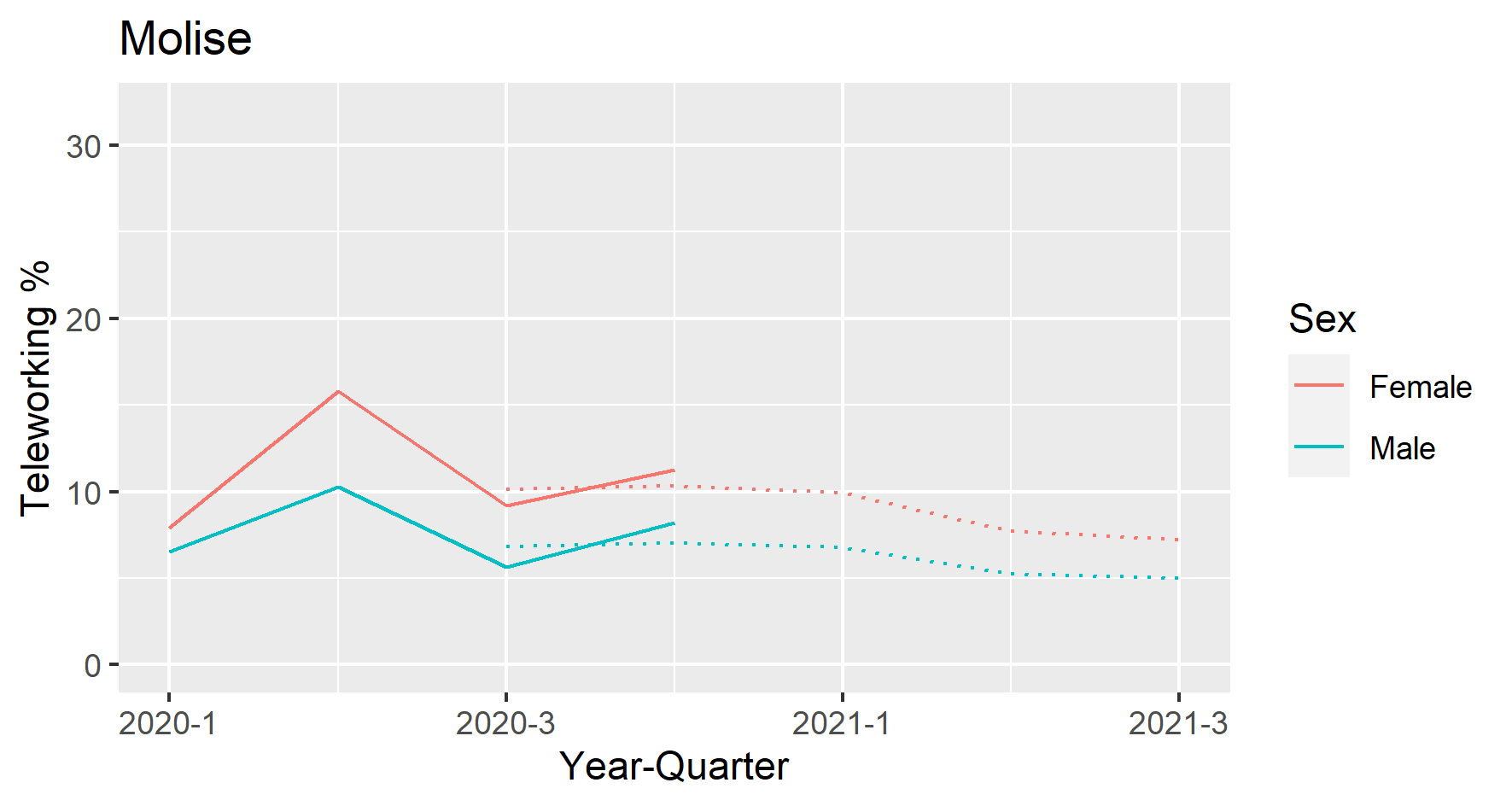}
\caption{Teleworking rate in Molise,  Labour Force Survey teleworking estimates in solid lines, mobility-derived teleworking estimates in dotted lines}
\end{figure*}

\begin{figure*}[!h]
\includegraphics[width=1\textwidth]{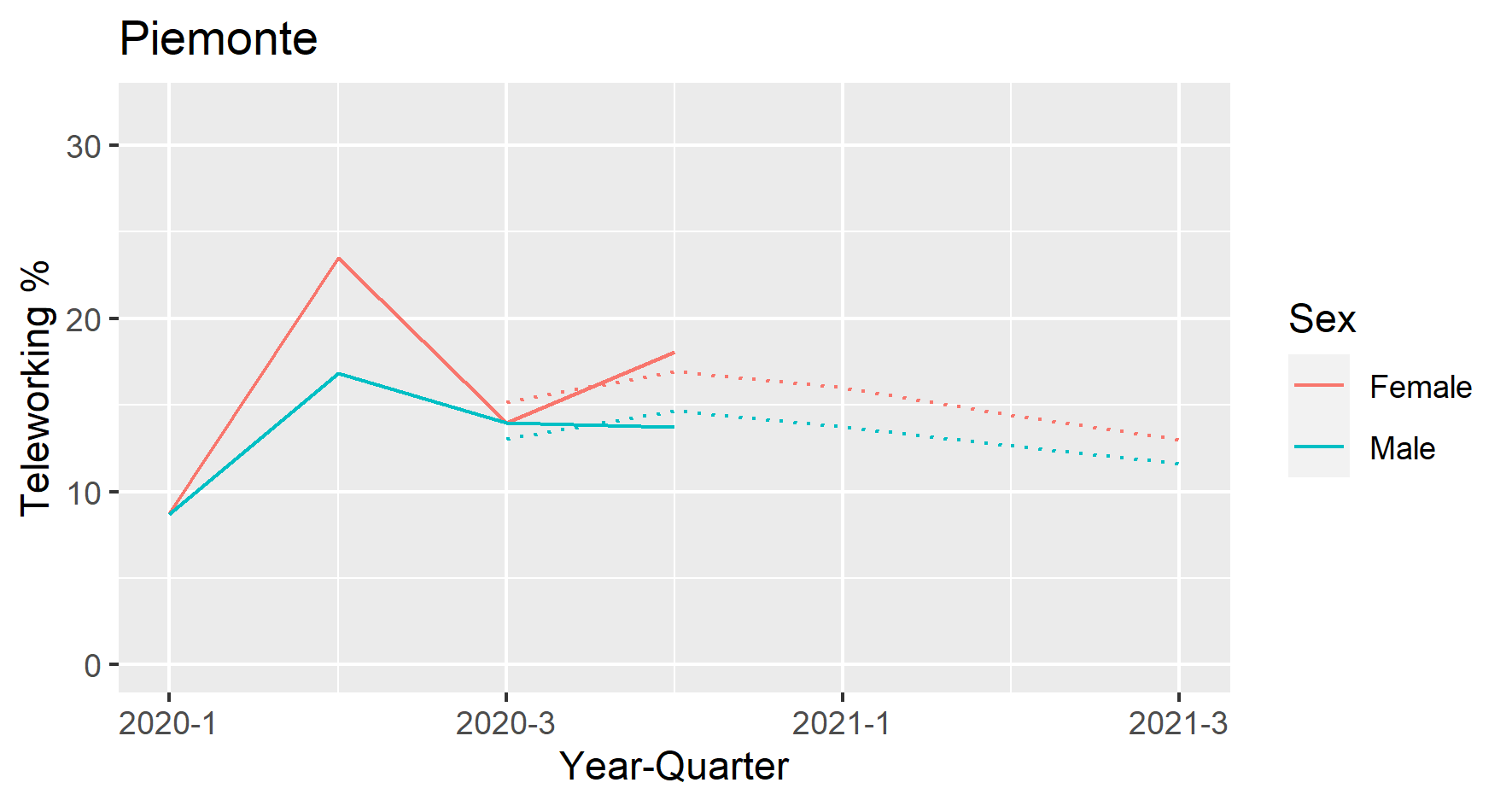}
\caption{Teleworking rate in Piemonte,  Labour Force Survey teleworking estimates in solid lines, mobility-derived teleworking estimates in dotted lines}
\end{figure*}

\begin{figure*}[!h]
\includegraphics[width=1\textwidth]{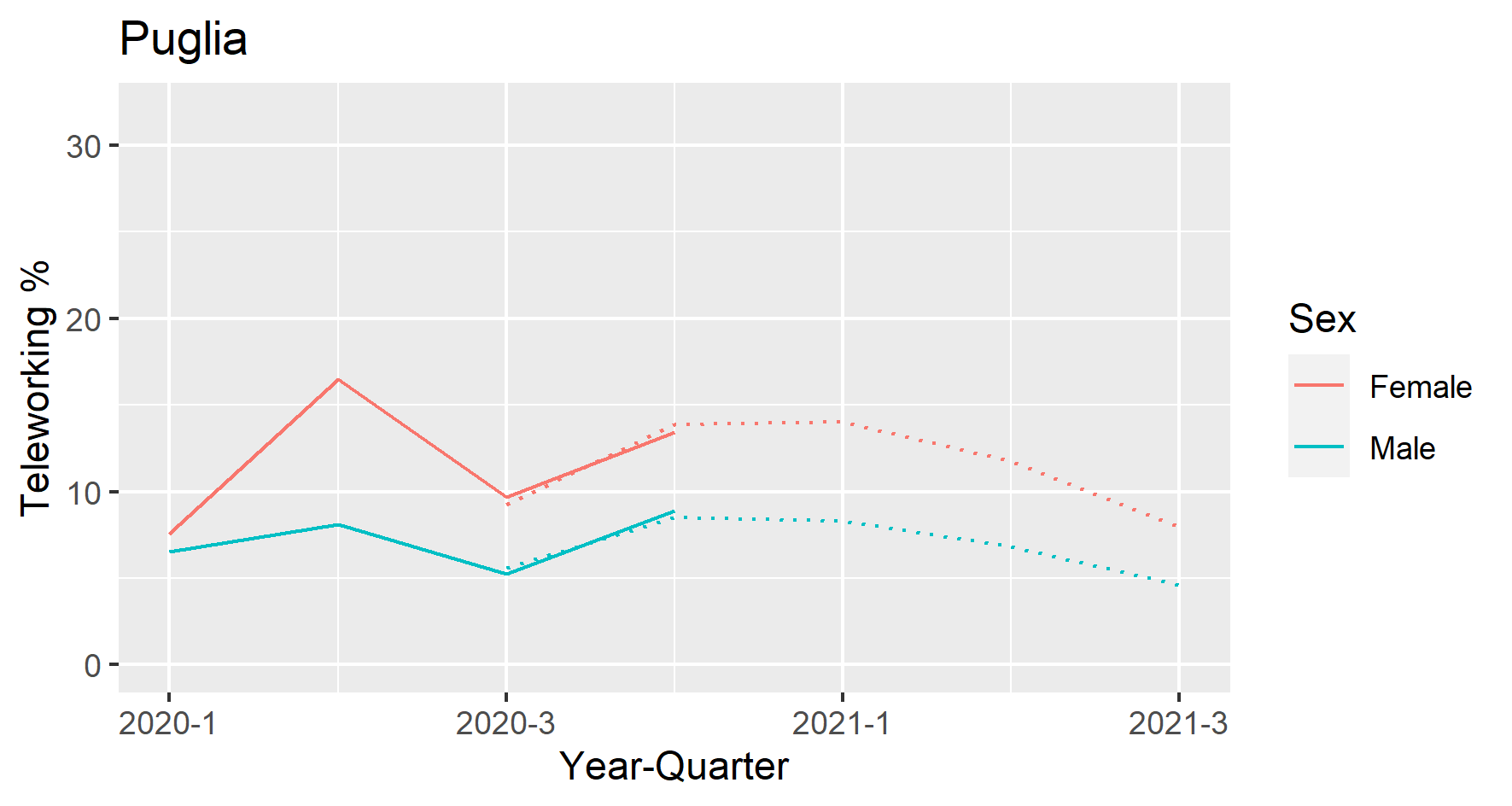}
\caption{Teleworking rate in Puglia,  Labour Force Survey teleworking estimates in solid lines, mobility-derived teleworking estimates in dotted lines}
\end{figure*}

\begin{figure*}[!h]
\includegraphics[width=1\textwidth]{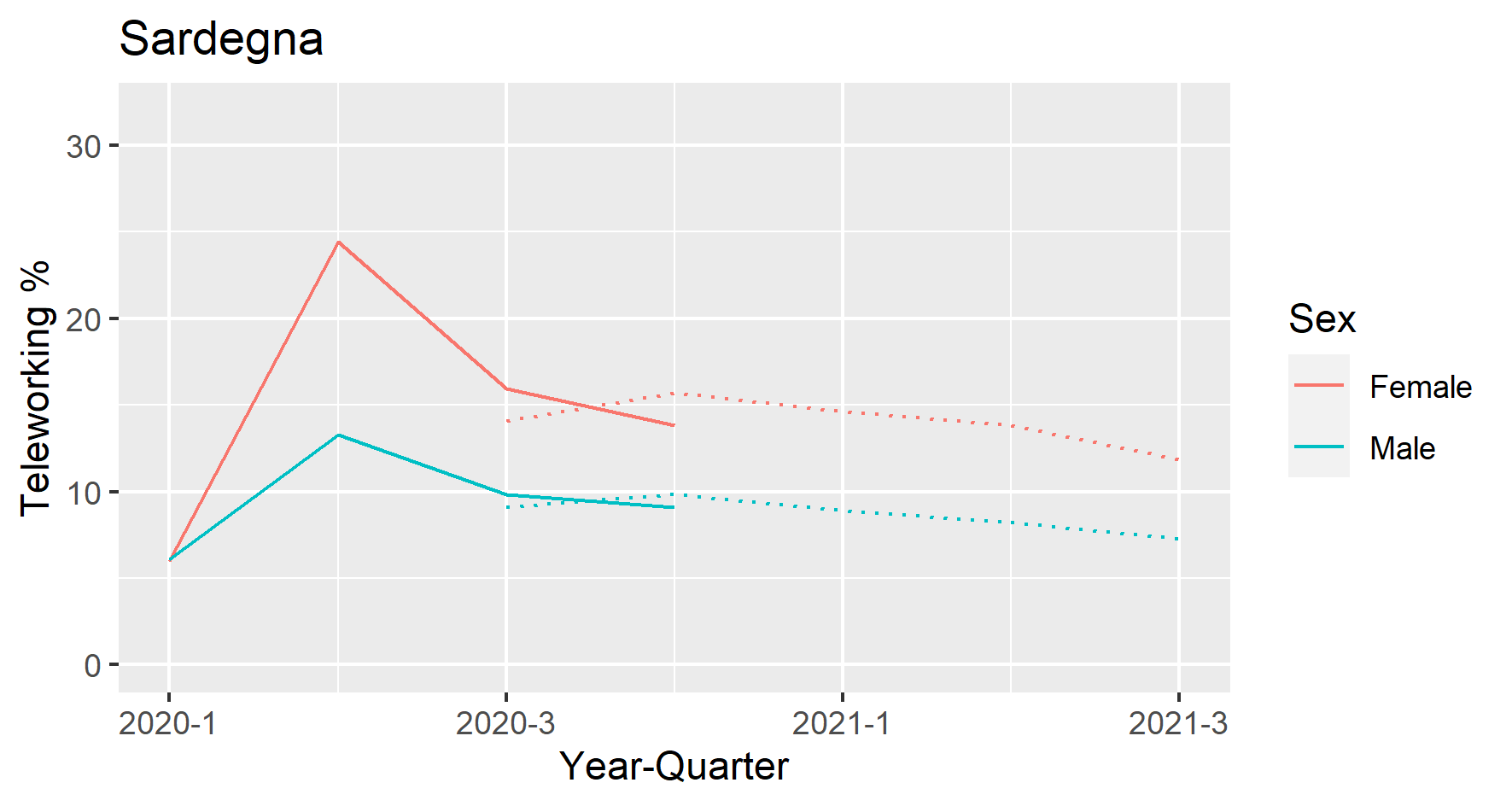}
\caption{Teleworking rate in Sardegna,  Labour Force Survey teleworking estimates in solid lines, mobility-derived teleworking estimates in dotted lines}
\end{figure*}

\begin{figure*}[!h]
\includegraphics[width=1\textwidth]{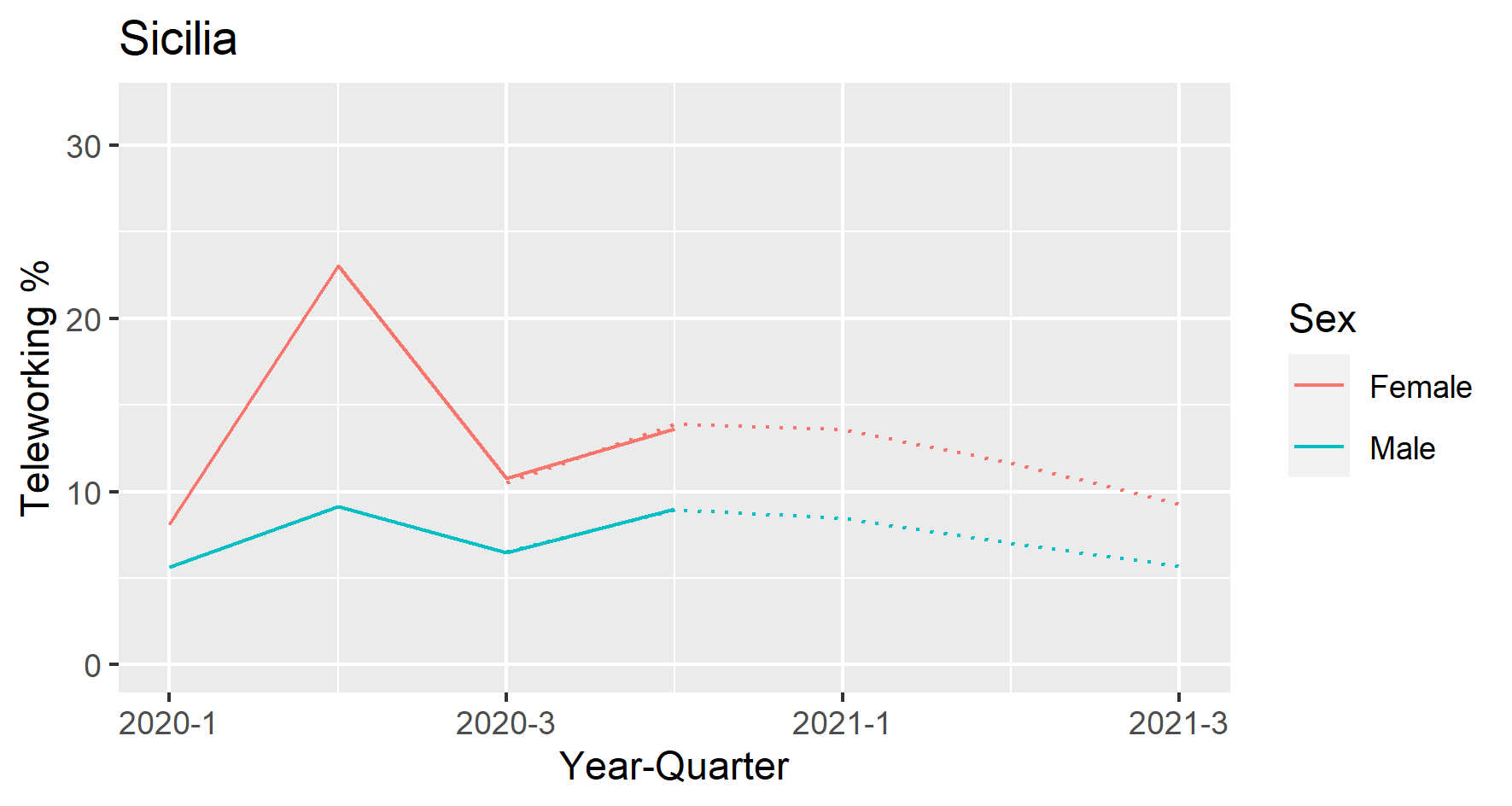}
\caption{Teleworking rate in Sicilia,  Labour Force Survey teleworking estimates in solid lines, mobility-derived teleworking estimates in dotted lines}
\end{figure*}

\begin{figure*}[!h]
\includegraphics[width=1\textwidth]{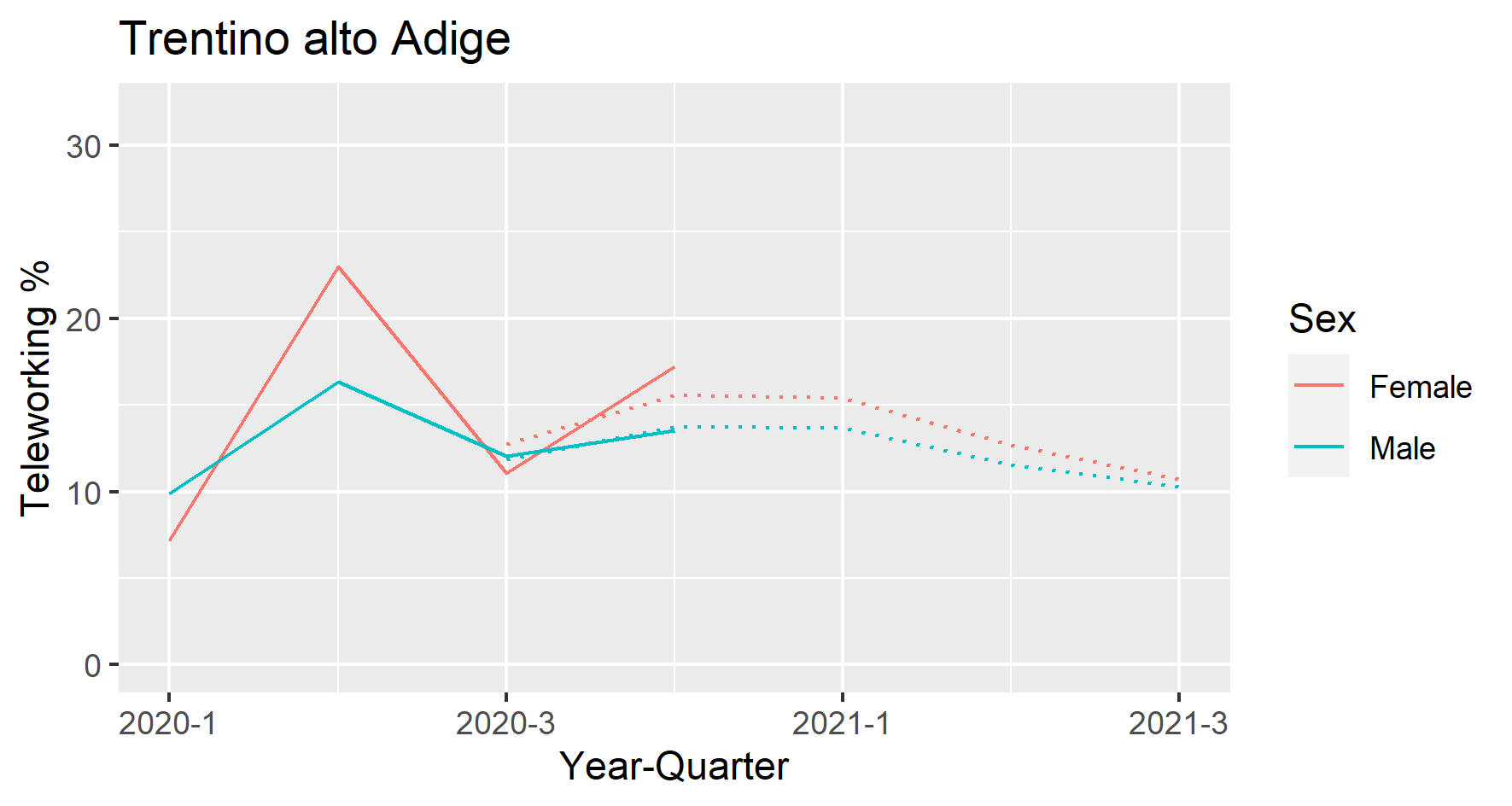}
\caption{Teleworking rate in Trentino alto Adige,  Labour Force Survey teleworking estimates in solid lines, mobility-derived teleworking estimates in dotted lines}
\end{figure*}

\begin{figure*}[!h]
\includegraphics[width=1\textwidth]{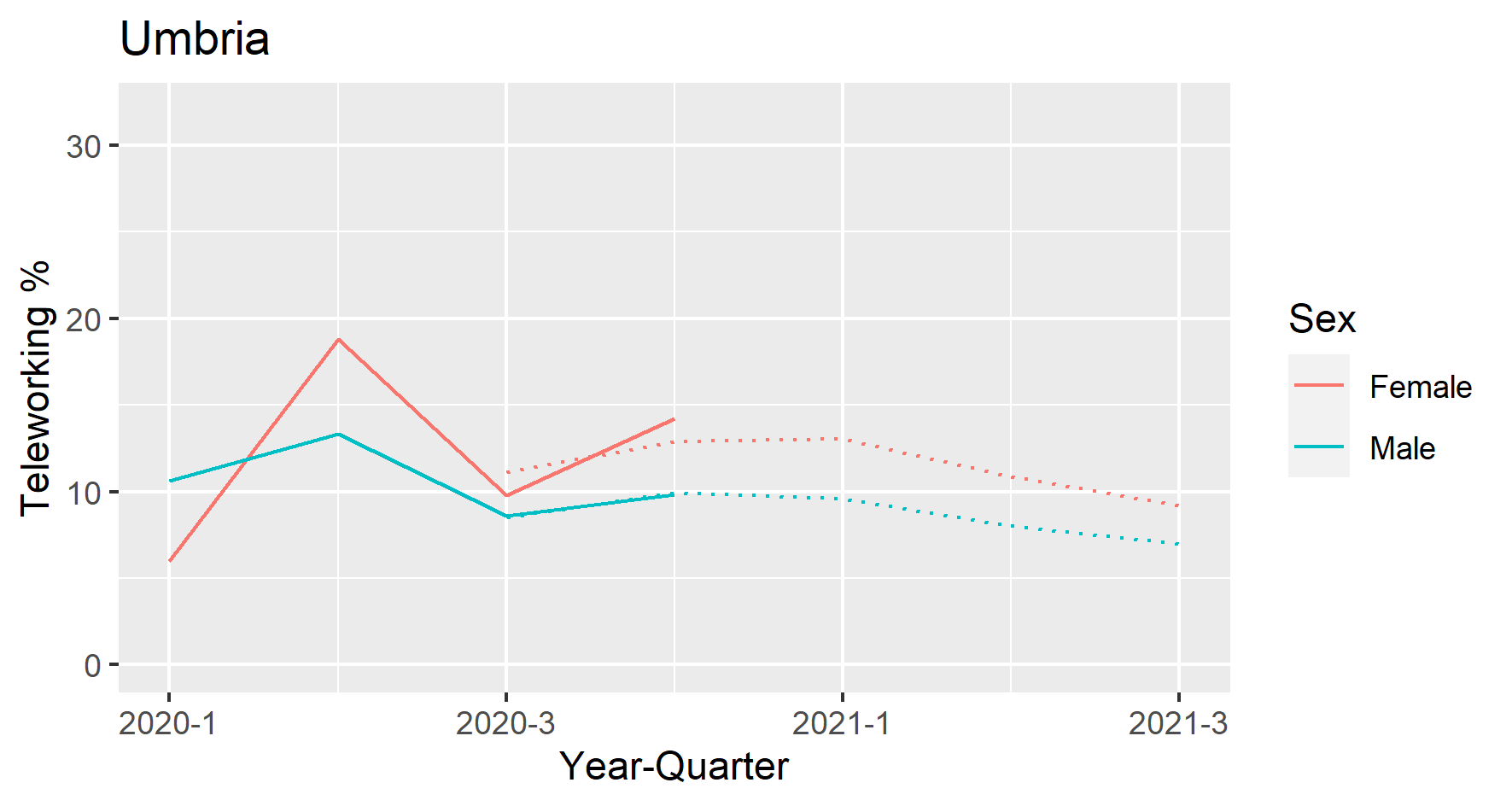}
\caption{Teleworking rate in Umbria,  Labour Force Survey teleworking estimates in solid lines, mobility-derived teleworking estimates in dotted lines}
\end{figure*}

\begin{figure*}[!h]
\includegraphics[width=1\textwidth]{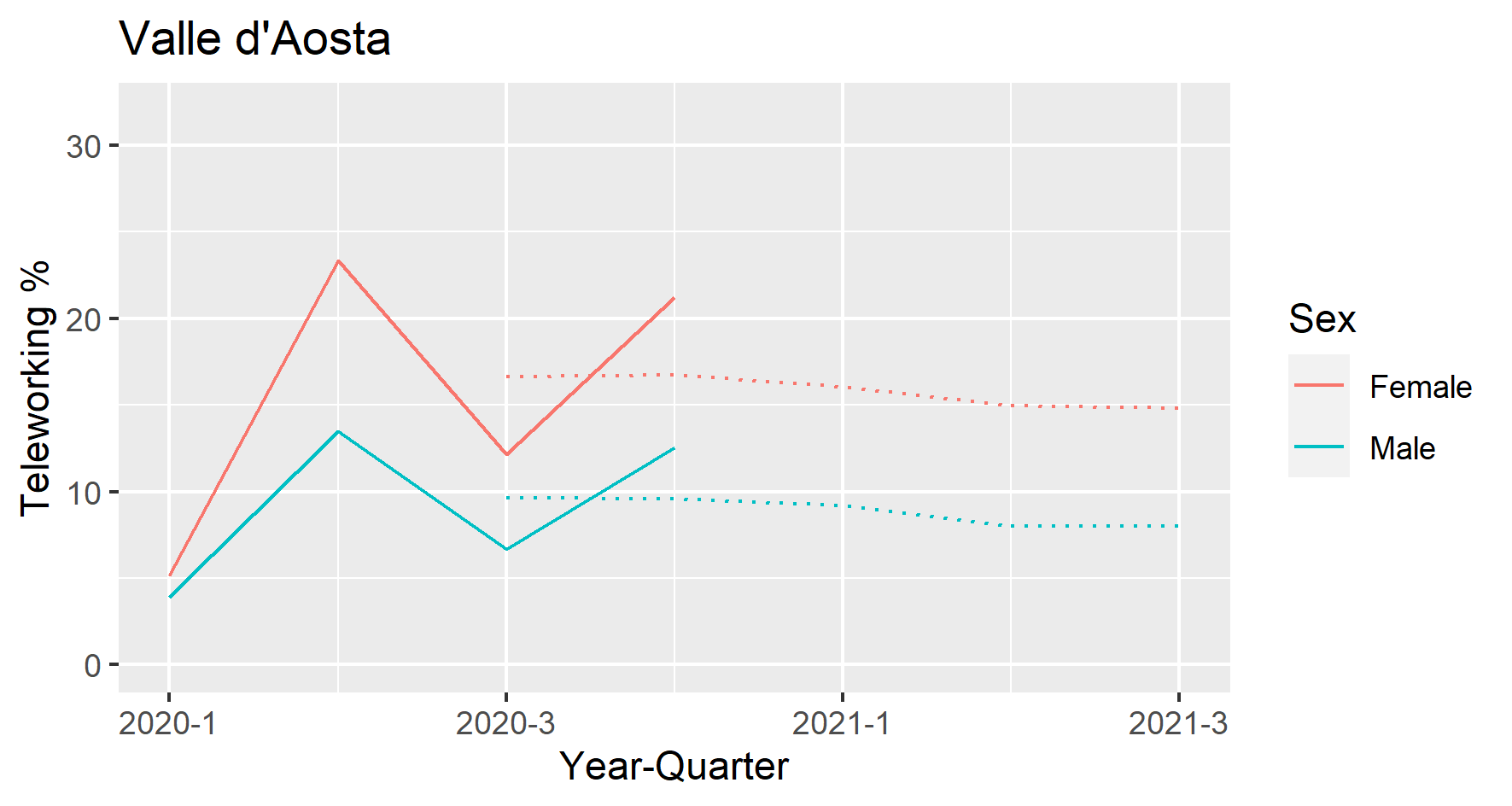}
\caption{Teleworking rate in Valle d'Aosta,  Labour Force Survey teleworking estimates in solid lines, mobility-derived teleworking estimates in dotted lines}
\end{figure*}

\begin{figure*}[h]
\includegraphics[width=1\textwidth]{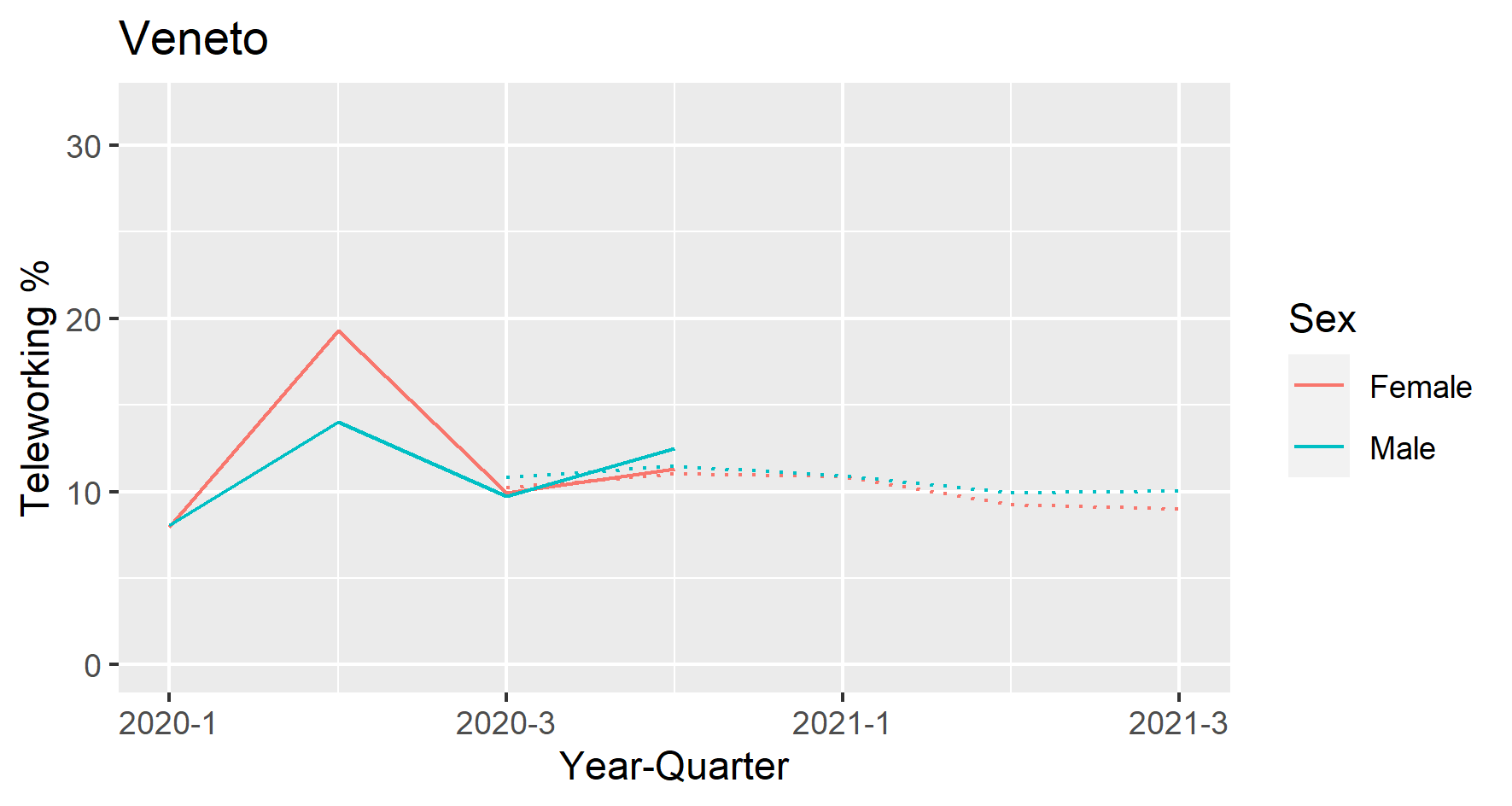}
\caption{Teleworking rate in Veneto,  Labour Force Survey teleworking estimates in solid lines, mobility-derived teleworking estimates in dotted lines}
\end{figure*}

\end{appendices}

\end{document}

%% file: model_minimal.tex
\begin{table}[!htbp] \centering 
  \caption{Regression output and descriptive statistics of the model described in equation \ref{eq:mod1}} 
  \label{minimal model} 

\begin{tabular}{@{\extracolsep{5pt}}lcc} 
\\[-1.8ex]\hline 
\hline \\[-1.8ex] 
 & \multicolumn{2}{c}{\textit{Dependent variable:}} \\ 
\cline{2-3} 
\\[-1.8ex] & \multicolumn{2}{c}{\textit{Teleworking}} \\ 
\hline \\[-1.8ex] 
employment & \multicolumn{2}{c}{0.041$^{*}$ (0.021)} \\ 
log(mobility) & \multicolumn{2}{c}{$-$0.169$^{***}$ (0.020)}\\ 
Constant & \multicolumn{2}{c}{$-$0.091$^{***}$ (0.031)} \\ 
\cline{2-3} 
\\[-1.8ex] & Male & Female \\ 
\cline{2-3} 
Abruzzo & $-$0.011 (0.017)  & $-$0.029 (0.019) \\ 
 Basilicata & $-$0.053$^{***}$ (0.018) & $-$0.034$^{*}$ (0.020) \\ 
 Calabria & $-$0.115$^{***}$ (0.019) & $-$0.049$^{**}$ (0.022) \\ 
 Campania & $-$0.045$^{**}$ (0.018) & 0.004 (0.021) \\ 
 Emilia Romagna & $-$0.018 (0.018) & $-$0.048$^{**}$ (0.020) \\ 
 Friuli Venezia Giulia & $-$0.007 (0.017) & 0.002 (0.019) \\ 
 Lazio & 0.072$^{***}$ (0.017)  & 0.094$^{***}$ (0.018) \\ 
 Liguria & $-$0.021 (0.019) & $-$0.043$^{**}$ (0.021) \\ 
 Lombardia & 0.076$^{***}$ (0.018)  & 0.083$^{***}$ (0.018) \\ 
 Marche & $-$0.047$^{***}$ (0.017) & $-$0.044$^{**}$ (0.019) \\ 
 Molise & $-$0.033$^{*}$ (0.018) & $-$0.014 (0.019) \\ 
 Piemonte & 0.020 (0.017)  & 0.012 (0.019) \\ 
 Puglia  & $-$0.130$^{***}$ (0.021)  & $-$0.124$^{***}$ (0.026) \\ 
 Sardegna & $-$0.045$^{**}$ (0.018) & $-$0.023 (0.020) \\ 
 Sicilia & $-$0.118$^{***}$ (0.021)  & $-$0.107$^{***}$ (0.025) \\ 
 Toscana  & $-$0.015 (0.018) & $-$0.027 (0.020) \\ 
 Trentino alto Adige & 0.009 (0.017) & 0.0002 (0.018) \\ 
 Umbria & $-$0.059$^{***}$ (0.018)  & $-$0.057$^{***}$ (0.020) \\ 
 Valle d'Aosta & 0.022 (0.018) & 0.073$^{***}$ (0.018) \\ 
 Veneto & & $-$0.031$^{*}$ (0.018) \\ 
 \hline \\[-1.8ex] 
Observations & \multicolumn{2}{c}{320} \\ 
R$^{2}$ & \multicolumn{2}{c}{0.606} \\ 
Adjusted R$^{2}$ & \multicolumn{2}{c}{0.548} \\ 
Residual Std. Error & \multicolumn{2}{c}{0.035 (df = 278)} \\ 
F Statistic & \multicolumn{2}{c}{10.422$^{***}$ (df = 41; 278)} \\ 
\hline 
\hline \\[-1.8ex] 
\textit{Note:}  & \multicolumn{2}{r}{$^{*}$p$<$0.1; $^{**}$p$<$0.05; $^{***}$p$<$0.01} \\ 
\end{tabular} 

\end{table} 